\documentclass[aps,pra,showpacs,floatfix,superscriptaddress]{revtex4}
\usepackage{graphicx}
\usepackage{color}
\usepackage{amssymb}
\usepackage{subfigure}
\usepackage{ulem}

\newtheorem{thm}{Theorem}

\begin{document}
\title{Complementarity in generic open quantum systems}
\author{Subhashish Banerjee}
\email{subhashish@cmi.ac.in}
\affiliation{Chennai Mathematical Institute, Padur PO, Siruseri 603103, India}
\affiliation{Raman Research Institute, Sadashiva Nagar,
Bangalore - 560 080, India}
\author{R. Srikanth}
\email{srik@ppisr.res.in}
\affiliation{Poornaprajna Institute of Scientific Research, Sadashiva
Nagar, Bangalore- 560 080, India.}
\affiliation{Raman Research Institute, Sadashiva Nagar,
Bangalore - 560 080, India}

\begin{abstract}
  We develop  a unified,  information theoretic interpretation  of the
  number-phase   complementarity   that    is   applicable   both   to
  finite-dimensional  (atomic)  and infinite-dimensional  (oscillator)
  systems, with number treated  as a discrete Hermitian observable and
  phase as a continuous  positive operator valued measure (POVM).  The
  relevant uncertainty principle is obtained  as a lower bound on {\it
    entropy excess},  $X$, the difference  between the entropy  of one
  variable,   typically  the   number,  and   the  knowledge   of  its
  complementary variable,  typically the  phase, where knowledge  of a
  variable  is defined  as its  relative entropy  with respect  to the
  uniform distribution.  In the  case of finite dimensional systems, a
  weighting of phase knowledge by  a factor $\mu$ ($> 1$) is necessary
  in order to make the bound tight, essentially on account of the POVM
  nature of phase as  defined here.  Numerical and analytical evidence
  suggests that $\mu$ tends to 1 as system dimension becomes infinite.
  We  study the  effect of  non-dissipative and  dissipative  noise on
  these  complementary  variables for  oscillator  as  well as  atomic
  systems.
\end{abstract}

\maketitle

\section{Introduction}

Two  observables  $A$  and  $B$  of  a  $d$-level  system  are  called
complementary  if  knowledge of  the  measured  value  of $A$  implies
maximal  uncertainty of  the measured  value  of $B$,  and vice  versa
\cite{mu88}.   Complementarity is  an aspect  of  the Heisenberg
uncertainty  principle, which  says  that for  any  state $\psi$,  the
probability  distributions obtained  by measuring  $A$ and  $B$ cannot
both  be   arbitrarily  peaked  if   $A$  and  $B$   are  sufficiently
non-commuting.  Expressed in terms of measurement entropy 
the Heisenberg  uncertainty principle takes the form:
\begin{equation}
H(A) + H(B) \ge \log d.
\label{eq:hu0}
\end{equation}
where $H(A)$ and $H(B)$ are the Shannon entropy of the
measurement outcomes of a $d$-level quantum system \cite{nc00,delg,op93}. Eq.
(\ref{eq:hu0}) has several advantages over the traditional 
uncertainty multiplicative form \cite{kraus,mu88,deu83,par83}.

More generally,
given two  observables $A \equiv  \sum_a a|a\rangle\langle a|$  and $B
\equiv  \sum_b b|b\rangle\langle  b|$,  let the  entropy generated  by
measuring  $A$  or  $B$  on   a  state  $|\psi\rangle$  be  given  by,
respectively,   $H(A)$   and   $H(B)$.   The   information   theoretic
representation  of the  Heisenberg uncertainty  principle  states that
$H(A) + H(B)  \ge 2\log\left(\frac{1}{f(A,B)}\right)$, where $f(A,B) =
\max_{a,b}|\langle a|b\rangle|$, and  $H(\cdot)$ is the Shannon binary
entropy.   
A pair of observables, $A$ and $B$,
for which  $f(A,B)=d^{-1/2}$ are said to form  mutually unbiased bases
(MUB) \cite{ii81,dur05}.  
Conventionally,  two Hermitian  observables  are called  complementary
only if they are mutually unbiased.  

An application of this idea to obtain an entropic uncertainty relation
for oscillator systems in the Pegg-Barnett scheme \cite{pb89} has been
made in Ref. \cite{abe}, and for higher entropic uncertainty relations
in  Ref.  \cite{wiwe}.  An  algebraic  treatment  of  the  uncertainty
relations,  in  terms  of  complementary subalgebras,  is  studied  in
Ref. \cite{petz}.

An extension of Eq. (\ref{eq:hu0}) to the case where $A$ or $B$ is not
discrete is considered in Ref. \cite{rs07}, where the problem that the
Shannon entropy  of a  continuous random variable  may be  negative is
circumvented   by  instead   using  relative   entropy   (also  called
Kullb{\"a}ck-Leibler divergence, which is always positive) \cite{kl51,
  hj03} with  respect to a  uniform distribution.  This quantity  is a
measure  of knowledge  \cite{rs07}.  An  example of  where  this finds
application would be when one of the observables, say $A$, is bounded,
and its conjugate $B$ is described  not as a Hermitian operator but as
a continuous-valued POVM.  A  particular case of this kind, considered
in detail  in Ref. \cite{rs07}, is  the number and phase  of an atomic
system.  This generalization of  the entropic uncertainty principle to
cover discrete-continuous  systems still suffers  from the restriction
that the  system must be finite  dimensional, since in the  case of an
infinite-dimensional system, such as an oscillator, entropic knowledge
of  the number  distribution  can diverge,  making  it unsuitable  for
infinite-dimensional systems.  Therefore to set up an entropic version
of the  uncertainty principle, that  unifies and is applicable  to all
systems,  including  infinite  dimensional and/or  continuous-variable
systems, it  may be advantageous to  use a combination  of entropy and
knowledge,  in  particular,  the  difference between  entropy  of  the
discrete, infinite  observable and  between phase knowledge.   This is
discussed in detail below.

The theory of  open quantum systems addresses the  problems of damping
and  dephasing in  quantum  systems  by its  assertion  that all  real
systems of interest are in fact `open' systems, each surrounded by its
environment.  One of  the first testing grounds for  open system ideas
was   in    quantum   optics   \cite{wl73}.     Depending   upon   the
system-reservoir  ($S-R$)  interaction, open  systems  can be  broadly
classified  into two categories,  viz., quantum  non-demolition (QND),
where  $[H_S,   H_{SR}]  =  0$  resulting  in   pure  decoherence,  or
dissipative,  where $[H_S,  H_{SR}] \neq  0$ resulting  in decoherence
along with dissipation \cite{bg06}.

The  plan of  the paper  is  as follows.   In Section  II, we  briefly
introduce, in anticipation of the discussion to follow, the concept of
quantum phase distributions for oscillator as well as two-level atomic
systems.   In  Section  III,   we  develop  an  information  theoretic
representation  of complementarity.   A  central feature  here is  the
study of  number-phase complementarity using the  principle concept of
{\it entropy excess}, the  difference between number entropy and phase
knowledge, mentioned above.   The use of the entropy  excess enables a
unified,  information  theoretic  interpretation of  the  number-phase
complementarity,  with  dimension-independent  lower  bound,  that  is
applicable     both     to     finite-dimensional     (atomic)     and
infinite-dimensional  (oscillator)   systems,  as  well   as  discrete
(number) and continuous (phase) variables.

We  apply  this entropic  uncertainty  principle  to various  physical
systems: oscillator systems (both  harmonic as well as anharmonic), in
Section IV, and atomic systems, in Section V, for a host of physically
relevant  initial  conditions.   In  addition, the  effect  of  purely
dephasing as well as dissipative influences on the system's evolution,
due to interaction with its  environment, and hence the entropy excess
is studied for each case considered  in Sections IV and V.  In Section
VI, we make our conclusions.

\section{Quantum Phase Distributions}

The  quantum description  of  phases \cite{pp98}  has  a long  history
\cite{pb89,pad27,   sg64,   cn68,ssw90,   scr93};   see   also   Refs.
\cite{ssw91, mh91}.  In a recent approach, which we adopt, the concept
of  phase  distribution for  the  quantum  phase  has been  introduced
\cite{ssw91,as92}.   Here we  briefly  recapitulate, for  convenience,
some  useful formulas  of quantum  phase distributions  for oscillator
systems \cite{sb06,sr07}.   For the case of atomic  systems, the basic
formulas were presented in \cite{rs07}.
 
Following  Agarwal  {\it  et  al.}   \cite{as92}  we  define  a  phase
distribution ${\cal  P}(\theta)$ for a given  density operator $\rho$,
which in our case would be the reduced density matrix, as
\begin{eqnarray}
{\cal P}(\theta) &=& {1 \over 2\pi} \langle \theta|\rho| \theta 
\rangle, ~ 0 \leq \theta \leq 2\pi, \nonumber\\ &=& {1 \over 
2\pi} \sum\limits_{m, n=0}^{\infty} \rho_{m, n} e^{ i(n-
m)\theta},  \label{osph} 
\end{eqnarray}
where  the   states  $|\theta\rangle$  are  the   eigenstates  of  the
Susskind-Glogower   \cite{sg64}   phase   operator  corresponding   to
eigenvalues of unit  magnitude and are defined in  terms of the number
states $|n\rangle$ as
\begin{equation} 
|\theta\rangle = \sum\limits_{n=0}^{\infty} e^{i n \theta}.
|n\rangle, \label{sg} 
\end{equation} 
The  sum in  Eq.   (\ref{osph})  is assumed  to  converge.  The  phase
distribution is positive definite and normalized to unity with
$\int_\theta |\theta\rangle\langle\theta|d\theta=1$.

The complementary number distribution is
\begin{equation}
p(m) = \langle m|\rho| m \rangle, \label{osnu}
\end{equation}
where $|m \rangle$ is the number (Fock) state.
Analogous results exist for atomic states, with the Susskind-Glagower
states replaced by atomic  coherent  states
\cite{mr78, ap90}, and number states by Wigner-Dicke states
\cite{at72}.

\section{Information theoretic representation of complementarity
\label{sec:qinf}}

Defining entropic knowledge $R[f]$ of random variable $f$ as its relative
entropy with
respect  to the uniform  distribution $\frac{1}{d}$, i.e.,
\begin{equation}
R[f] \equiv S\left(f(j)||\frac{1}{d}\right) = \sum_j f(j)\log(df(j)),
\label{eq:rf}
\end{equation}
we can recast Heisenberg  uncertainty  principle in
terms  of entropy  $H$ and  knowledge $R$, as shown by this easy theorem
\begin{thm}
Given  two  Hermitian  observables  $A$  and  $B$ that form a
pair of MUB,  the
uncertainty relation (\ref{eq:hu0}) can be expressed as
\begin{equation}
X(A,B) \equiv H(A) - R(B) \ge 0.
\label{eq:ra}
\end{equation}
\end{thm}
{\bf Proof.} Let the distribution obtained by measuring $A$ and $B$ on
a given state be, respectively, $\{p_j\}$ and $\{q_k\}$.  Denoting
$H(A) \equiv \sum_j p_j \log_2 p_j$, the l.h.s of Eq. (\ref{eq:ra}) is
given by
\begin{eqnarray}
H\left(A\right) - S\left(B||\frac{1}{d}\right) &=&
H(A) - \sum_k q_k \log( dq_k) \nonumber \\
&=& H(A) + H(B) - \log d \label{eq:sub} \\
&\ge & 2\log\left(\frac{1}{f(A,B)}\right) - \log d \label{eq:sri1}.
\end{eqnarray}
where Eq. (\ref{eq:sri1}) follows from Ref. \cite{mu88}.
For   a   pair   of  MUB  \cite{kraus,deu83},
$f(A,B)=d^{-1/2}$, from which the theorem follows.  \hfill $\blacksquare$
\bigskip \\ 

From Eq.  (\ref{eq:sub}) it follows  that $X(A,B)=X(B,A)$.  Therefore,
phyically Eq. (\ref{eq:ra}) expresses that ignorance of one of two MUB
variables is  at least as large as  the knowledge of the  other. It is
not  difficult to  see  that  $X(A,B)$ attains  its  largest value  of
$\log(d)$  when  $A$  and $B$  are  MUBs,  and  its minimum  value  of
$-\log(d)$  when $A$  and  $B$ are  identical.   This gives  a way  to
quantify the  `degree of complementarity'.  Define $X_{\min}(A,B)$ are
the smallest  value of $X(A,B)$ over  all possible states  for a given
pair of Hermitian observables $A$  and $B$.  Then, two observables $A$
and $B$ are maximally  complementary (i.e., MUB) if $X_{\min}(A,B)=0$,
and   they   are   minimally   complementary  (i.e.,   identical)   if
$X_{\min}(A,B)=-\log(d)$.

A point  worth noting about Eq.  (\ref{eq:ra}) is that  it contains no
explicit mention of dimension $d$.  What is remarkable is that we find
this situation persists  even when one of $A$ or  $B$ is not discrete,
but  a continuous-valued  POVM  (for discrete-valued  POVMs, cf.  Ref.
\cite{mas07}),  and  furthermore,  the  system  is  no  longer  finite
dimensional  but  instead infinite  dimensional.  The only  additional
requirement is  that the continuous-valued  variable should be  set as
$B$ (the knowledge- rather  than the ignorance-variable), since $H(B)$
can potentially  be negative for  such variables.  This  makes $X(A,B)
\ge 0$  as a  very succinct and  general statement of  the uncertainty
principle.   By contrast,  because there  is no  prior  guarantee that
measurement   entropy   $H(\cdot)$   will   be  non-negative   for   a
continuous-valued observable,  it is not  obvious that the  version of
the  Heisenberg  uncertainty  principle  given  by  (\ref{eq:hu0})  is
generally  applicable,  and furthermore,  because  there  is no  prior
guarantee  that  measurement  entropic  knowledge $R(\cdot)$  will  be
well-defined for  infinite-dimensional variables, the  version $R(A) +
R(B) \le \log(d)$ of Ref.  \cite{rs07} is also not obviously generally
applicable.

One catch  is that on  account of the  POVM-nature of $B$,  $R(B)$ may
have a  maximum value  less than $\log(d)$  in the  finite dimensional
case.   It   will  be   to   generalize   the   concept  of   `maximal
complementarity' or  `MUBness' to  apply those terms  to $A$  and $B$,
when one of  them is a POVM, if the maximal  knowledge of the measured
value of $A$  implies minimal knowledge of the  measured value of $B$,
and vice versa, but with maximum knowledge no longer being required to
as high as $\log d$ bits.

For the phase variable given by the POVM $\phi$ and probability
distribution ${\cal P}(\phi)$, entropic knowledge
is   given  by  the functional \cite{sb06,sr07}:
\begin{equation}
R[{\cal P}(\phi)] = \int_0^{2\pi}d\phi~
{\cal P}(\phi)\log[2\pi {\cal P}(\phi)],
\label{eq:phient}
\end{equation}
where the $\log(\cdot)$ refers to the binary base.

It is at  first not obvious that Eq.  (\ref{eq:ra}) holds for infinite
dimensional systems. Based on a result due to Ref.  \cite{bial} for an
oscillator  system,  which in turn uses the concept of the
$(p, q)$-norm of the Fourier transformation found by Beckner \cite{beck75} 
for all values of p, for an
oscillator system,  we  can show  that  it  is  indeed the  case.  In particular,
\begin{equation}
\label{eq:bial}
-\int_{-\pi}^\pi   d\phi   P(\phi)\log(P(\phi))  -   \sum_{m=0}^\infty
p_m\log(p_m) \ge \log(2\pi)
\end{equation}

Setting the `number  variable' $m$ in Eq. (\ref{eq:bial})  as $A$, and
the phase  variable $\phi$ as $B$,  and noting that the  first term in
the l.h.s of Eq. (\ref{eq:bial}), using Eq. (\ref{eq:phient}), is just
$\log(2\pi) - R[P(\phi)]$, we obtain
\begin{equation}
X[m, \phi] \equiv H[m] - R[\phi] \ge 0,
\label{eq:nybial}
\end{equation}
which is  Eq. (\ref{eq:ra}) applied to  an infinite-dimensional system
that   includes   a    non-Hermitian   POVM   (phase   $\phi$).    Eq.
(\ref{eq:nybial}) expresses  the fact ignorance of variable  $m$ is at
least  as great  as knowledge  of its  complementary  partner, $\phi$.
Comparing Eqs.  (\ref{eq:ra}) and  (\ref{eq:nybial}), we find that the
statement $X  \ge 0$  as a description  of the  Heisenberg uncertainty
relation holds good both  for finite and infinite dimensional systems.
The version $X  \ge 0$ of the Heisenberg  uncertainty principle may be
called  the principle  of  entropy excess.   An information  theoretic
interpretation of the above relation  has been studied, in the context
of phase  resolution in  harmonic oscillator systems,  in \cite{mh93}.
Also,  the  number-phase complementarity,  for  a harmonic  oscillator
system,  using information  exclusion  relations has  been studied  in
\cite{mh95}.

\section{Oscillator System}

Here we  consider the application  of the principle of  entropy excess
(\ref{eq:nybial})  to oscillator  systems,  both harmonic  as well  as
anharmonic,  starting  from  a   number  of  physically  relevant  and
interesting initial conditions  and interacting with their environment
via a purely  dephasing (QND) as well as  dissipative interaction. The
strategy would  be to compute  the phase and number  distributions for
each  case,  use them  to  obtain  phase knowledge  (\ref{eq:phient}),
number entropy  and use  them in Eq.   (\ref{eq:nybial}) to  study the
entropy excess and thus the number-phase complementarity in oscillator
systems.

\subsection{QND system-bath interaction}

Consider   the  following  Hamiltonian   describing  the
interaction of a system with  its environment, modelled as a reservoir
of harmonic oscillators, via a QND type of coupling :
\begin{eqnarray}
H & = & H_S + H_R + H_{SR} \nonumber\\ & = & H_S + 
\sum\limits_k \hbar \omega_k b^{\dagger}_k b_k + H_S 
\sum\limits_k g_k (b_k+b^{\dagger}_k) + H^2_S \sum\limits_k 
{g^2_k \over \hbar \omega_k}. \label{2a} 
\end{eqnarray} 
Here $H_S$, $H_R$ and $H_{SR}$ stand for the Hamiltonians of 
the system, reservoir and system-reservoir interaction, 
respectively. $H_S$ is a generic system Hamiltonian which we 
will specify in the subsequent sections to model different 
physical situations. $b^{\dagger}_k$, $b_k$ denote the creation 
and annihilation operators for the reservoir oscillator of 
frequency $\omega_k$, $g_k$ stands for the coupling constant 
(assumed real) for the interaction of the oscillator field with 
the system. The last term on the right-hand side of Eq. (1) is 
a renormalization inducing `counter term'. Since $[H_S, 
H_{SR}]=0$, the Hamiltonian (1) is of QND type. The system plus 
reservoir composite is closed obeying a unitary evolution given 
by 
\begin{equation}
\rho (t) = e^{-{i \over \hbar}Ht} \rho (0) e^{{i \over 
\hbar}Ht} , \label{2b} 
\end{equation}
where
\begin{equation}
\rho (0) = \rho^s (0) \rho_R (0),\label{2c}
\end{equation}
i.e., we assume separable initial conditions. The reservoir is 
assumed to be initially in a squeezed thermal state, i.e., it 
is a squeezed thermal bath, with an initial density matrix 
$\rho_R (0)$ given by 
\begin{equation}
\hat{\rho}_R(0) = \hat{S} (r,\Phi) \hat{\rho}_{th} 
\hat{S}^{\dagger} (r,\Phi),\label{2d} 
\end{equation}
where
\begin{equation}
\hat{\rho}_{th} = \prod_k \left[ 1 - e^{-\beta \hbar \omega_k} 
\right] e^{-\beta \hbar \omega_k \hat{b}^{\dagger}_k  
\hat{b}_k} \label{2e} 
\end{equation}
is the density matrix of the thermal bath, and
\begin{equation}
\hat{S} (r_k, \Phi_k) = \exp \left[ r_k \left( {\hat{b}^2_k 
\over 2} e^{-i2\Phi_k} - {\hat{b}^{\dagger 2}_k \over 2} 
e^{i2\Phi_k} \right) \right] \label{2f} 
\end{equation}
is the squeezing operator with $r_k$, $\Phi_k$ being the 
squeezing parameters \cite{cs85}. We are interested in the 
reduced dynamics of the `open' system of interest $S$, which is 
obtained by tracing over the bath degrees of freedom. Using 
Eqs. (\ref{2a}), (\ref{2c}) in Eq. (\ref{2b}) and tracing over 
the bath variables, we obtain the reduced density matrix for 
$S$, in the system eigenbasis, as \cite{bg06} 
\begin{eqnarray}
\rho^s_{nm} (t) & = & e^{-{i \over \hbar}(E_n-E_m)t} e^{
i(E^2_n-E^2_m)\eta(t)}
\times \exp \Big[-(E_m-E_n)^2 \gamma(t) \Big] 
\rho^s_{nm} (0). \label{qndred} 
\end{eqnarray}
In the  above equation, $E_n$ is  the eigenvalue of the  system in the
system eigenbasis while $\eta(t)$ and $\gamma (t)$ quantify the effect
of the  bath on the system  and are given  in Appendix \ref{secap:qnd}
for convenience.

\subsubsection{System of a harmonic oscillator}

\begin{figure}
\subfigure[]{\includegraphics[width=7.0cm]{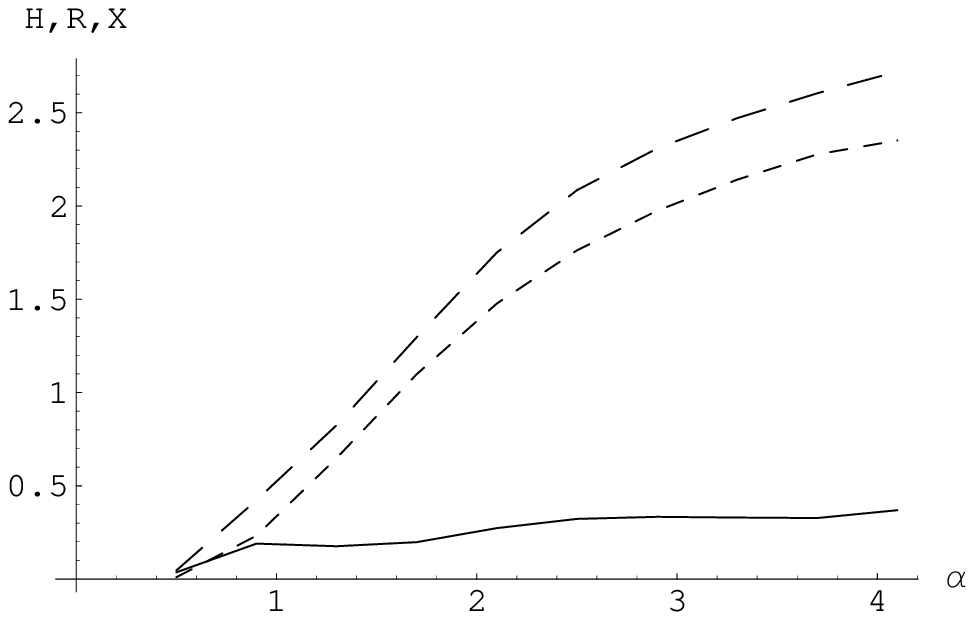}}
\subfigure[]{\includegraphics[width=7.0cm]{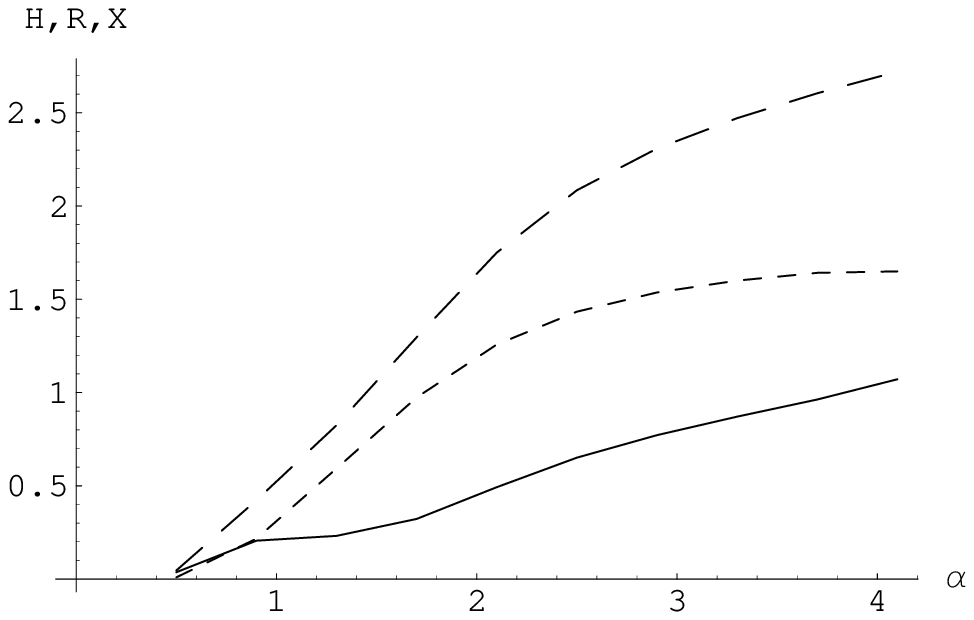}}
\caption{ Number  entropy $H[m]$ (large-dashed  line), phase knowledge
  $R[\theta]$  (small-dashed line) and  entropy excess  $X[m, \theta]$
  (Eq.   (\ref{eq:nybial}), bold line)  plotted as  a function  of the
  parameter  $\alpha$ (\ref{coh}) for  the harmonic  oscillator system
  initially in  a coherent state.   Figure (a) represents the  case of
  the  pure  state case.   We  note  that  as number  increases,  with
  increase  in $\alpha$,  so  does  $H[m]$ (since  the  variance of  a
  Poisson distribution equals its  mean), whereas phase $\phi$ becomes
  increasingly certain, leading to  increase in $R[\phi]$.  Figure (b)
  represents the case of the  system subjected to QND interaction with
  the    parameters     $\omega=1.0$,    $\omega_c=100$,    $\gamma_0$
  (\ref{gamma0})  $=0.0025$, $|\alpha|^2=5$,  $\theta_0=0$ (\ref{coh})
  and with  bath squeezing  parameters (\ref{eq:a}) $r=2.0$  and $a=0$
  for a temperature  $T$ (in units where $\hbar  \equiv k_B \equiv 1$)
  =1 and an evolution time $t=0.5$.}
\label{fig:inicoh}
\end{figure}

We  consider  the  system  $S$  of  a  harmonic  oscillator  with  the
Hamiltonian
\begin{equation} 
H_S = \hbar \omega \left(a^{\dag}a + {1 \over 2} \right). 
\label{harmonic} 
\end{equation}
The number states serve as an appropriate basis for the system 
Hamiltonian and the system energy eigenvalue in this basis is 
\begin{equation}
E_{n}= \hbar \omega \left( n + {1 \over 2}\right) . \label{3d} 
\end{equation}
The harmonic oscillator system is  assumed to start from the following
physically interesting initial states:

(A). \underline{System initially in a coherent state}:

The initial density matrix of the system is 
\begin{equation}
\rho^s(0) = |\alpha \rangle \langle\alpha|, \label{hocoh} 
\end{equation}
where 
\begin{equation}
\alpha = |\alpha| e^{i \theta_0} \label{coh} 
\end{equation}
is a coherent state  \cite{sz97}.  Making use of Eqs.  (\ref{qndred}),
(\ref{hocoh}) in Eq. (\ref{osph}),  the phase distribution is obtained
as \cite{sb06}
\begin{eqnarray}
{\cal P}(\theta) & = & {1 \over 2\pi} \sum\limits_{m, 
n=0}^{\infty} {|\alpha|^{n + m} \over \sqrt{n!m!}} e^{-
|\alpha|^2} e^{-i (m - n) (\theta - \theta_0)} e^{-i \omega(m - 
n)t} \nonumber \\ & & \times e^{i (\hbar \omega)^2 (m -n)(n + m 
+ 1) \eta(t)} e^{-(\hbar \omega)^2 (n - m)^2 \gamma(t)}. 
\label{cohosph} 
\end{eqnarray} 
The corresponding complementary number distribution is obtained, using
Eq. (\ref{osnu}), as
\begin{equation}
p(m) = \frac{|\alpha|^{2m}}{m!} e^{-|\alpha|^2}. \label{cohosnu}
\end{equation}
Using  ${\cal   P}(\theta)$   (\ref{cohosph})  in   Eq.
(\ref{eq:phient}) to  get the phase  knowledge, $p(m)$ (\ref{cohosnu})
to get the number entropy  and using these in Eq. (\ref{eq:nybial}) we
get   the   entropy   excess.     These   are   plotted   in   Figures
\ref{fig:inicoh}.   It is  clearly  seen, by  a  comparison of  Figure
\ref{fig:inicoh}(b)  with (a)  (representing unitary  evolution), that
including  the environmental  effects  due to  finite temperature  and
squeezing causes  the entropy excess to increase  by randomizing phase
and thus causing $R[\theta]$ to fall, whereas $H[m]$ remains invariant
because   QND  interactions   characteristically   leave  the   number
distribution  $p(m)$ (\ref{osnu}) invariant  \cite{sr07}. This  can be
seen from Eq.  (\ref{cohosnu}),  where the only parameter entering the
distribution  $p(m)$ is  the  initial state  parameter $\alpha$.   The
figures   clearly  show   that  the   principle  of   entropy  excess,
Eq. (\ref{eq:nybial}), is satisfied for both unitary evolution as well
as in the case of interaction with the bath. 

\begin{figure}
\subfigure[]{\includegraphics[width=7.0cm]{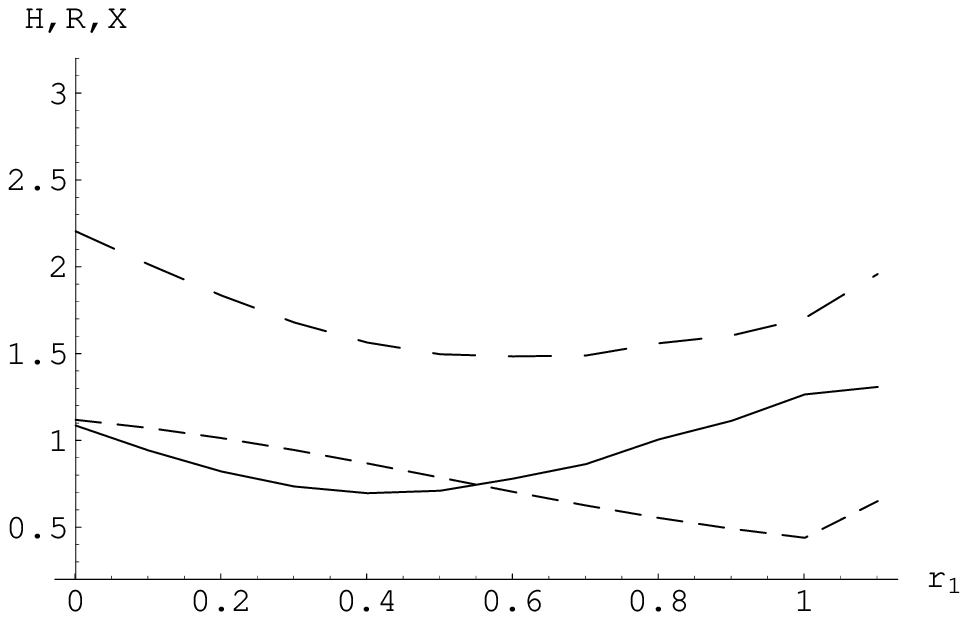}}
\subfigure[]{\includegraphics[width=7.0cm]{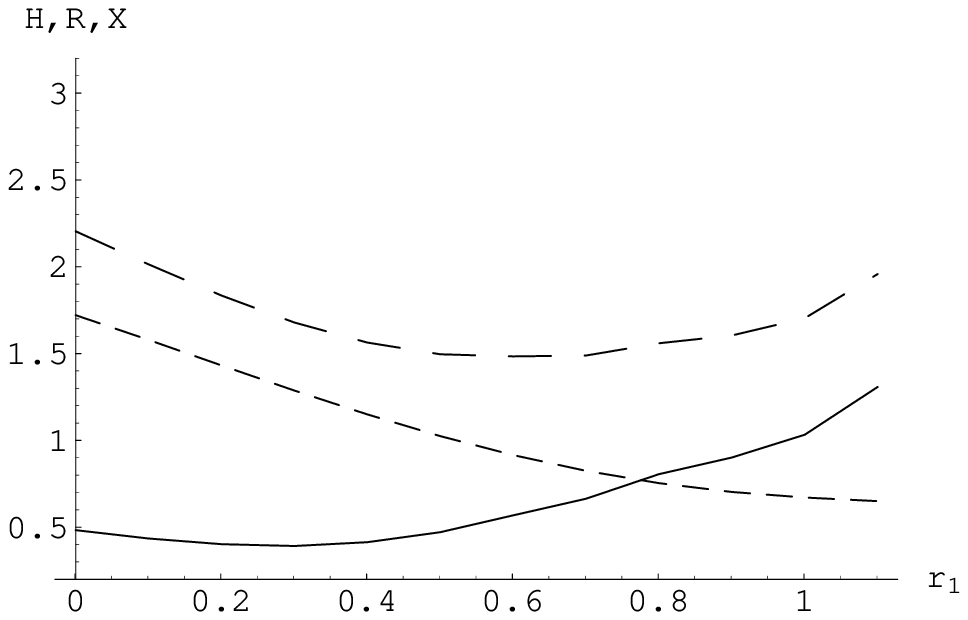}}
\caption{ Number  entropy $H[m]$  (large-dashed  line),
  phase knowledge  $R[\theta]$ (small-dashed line)  and entropy excess
  $X[m,  \theta]$ (Eq.   (\ref{eq:nybial})) (bold  line) plotted  as a
  function of  the system squeezing  parameter $r_1$ (\ref{sq})  for a
  harmonic oscillator  system initially  in a squeezed  coherent state
  and  subjected to  a  QND interaction,  with parameters  $\omega=1$,
  $\omega_c=100$,   $\gamma_0=0.025$,   $|\alpha|^2=5$,   $\theta_0=0$
  (\ref{coh}),  $\psi$   (\ref{sq})  $=0$  and   with  bath  squeezing
  parameters (\ref{eq:a}) $r=1.0$  and $a=0$. Figure (a)
  represents  an evolution time  $t=0.1$ and  $T=0$, while  figure (b)
  depicts the case for an evolution time $t=0.1$ and $T=1$.}
\label{fig:inisqcoh}
\end{figure}

(B). \underline{System initially in a squeezed coherent state}:

The initial density matrix of the system is 
\begin{equation}
\rho^s(0) = |\xi, \alpha \rangle \langle\alpha, \xi|, 
\label{hosq} 
\end{equation}
where the squeezed coherent state is defined as \cite{sz97} 
\begin{equation} 
|\xi, \alpha \rangle = S(\xi) D(\alpha) |0\rangle. \label{sq} 
\end{equation}
Here  $S$ denotes  the standard  squeezing operator with
$\xi  = r_1  e^{i \psi}$ and  $D$ denotes  the standard
displacement operator \cite{sz97}.  Making use of Eqs. (\ref{qndred}),
(\ref{hosq}) in  Eq. (\ref{osph}), the phase  distribution is obtained
as \cite{sb06}
\begin{eqnarray}
{\cal P}(\theta) &=& {1 \over 2\pi} \sum\limits_{m, 
n=0}^{\infty} e^{i(n-m)\theta} {e^{i{\psi \over 2}(m-n)} \over 
2^{(m+n) \over 2} \sqrt {m!n!}} {(\tanh(r_1))^{(m+n) \over 2} 
\over \cosh(r_1)} \nonumber \\ & & \times \exp \left[-
|\alpha|^2 (1 - \tanh(r_1) \cos(2\theta_0 - \psi)) \right] 
\nonumber \\ & & \times H_m \left[{|\alpha| e^{i(\theta_0 - 
{\psi \over 2})} \over \sqrt{\sinh(2r_1)}} \right] H^{*}_n 
\left[ {|\alpha| e^{i(\theta_0 - {\psi \over 2})} \over 
\sqrt{\sinh(2r_1)}} \right] \nonumber \\ & & \times  e^{-i 
\omega(m-n)t} e^{i (\hbar \omega)^2 (m-n)(n+m+1) \eta(t)} e^{-
(\hbar \omega)^2 (n-m)^2 \gamma(t)}. \label{sqosph} 
\end{eqnarray} 
Here   $H_n[z]$   is   a   Hermite  polynomial.    The   corresponding
complementary number distribution is obtained, using Eq. (\ref{osnu}),
as
\begin{equation}
p(m) = \frac{1}{2^m m!} \frac{(\tanh(r_1))^{m}}{\cosh(r_1)}
\exp \left[-
|\alpha|^2 (1 - \tanh(r_1) \cos(2\theta_0 - \psi)) \right]
|H_m \left[{|\alpha| e^{i(\theta_0 - 
{\psi \over 2})} \over \sqrt{\sinh(2r_1)}} \right]|^2. \label{sqosnu}
\end{equation}
Using   ${\cal   P}(\theta)$  (\ref{sqosph})   in   Eq.
(\ref{eq:phient}) to get the phase knowledge, $p(m)$ (\ref{sqosnu}) to
get the number entropy and using these in Eq. (\ref{eq:nybial}) we get
the entropy  excess which  are plotted in  Figures \ref{fig:inisqcoh}.
From the  Figures \ref{fig:inisqcoh}  it can be  seen that  phase gets
randomized, resulting  in a fall  in the phase  knowledge $R[\theta]$,
with increase in the system squeezing parameter $r_1$ (\ref{sq}).  The
number entropy $H[m]$ is not effected by the reservoir, due to the QND
nature of the interaction but as can be seen from Eq.  (\ref{sqosnu}),
the  number  distribution  $p(m)$   depends  upon  the  initial  state
parameters $\alpha$, $r_1$  and $\psi$.  Thus $H[m]$ as  a function of
the system squeezing  parameter $r_1$ first falls and  then rises as a
result of which the entropy excess  at first goes down and then rises.
The principle  of entropy  excess, Eq.  (\ref{eq:nybial}),  is clearly
seen to be satisfied.

An  interesting   feature   here  is   that  in   Figure
\ref{fig:inisqcoh}(b), even though in  comparison with the settings in
Figure \ref{fig:inisqcoh}(a) temperature  $T$ has increased, the value
of $R[\theta]$  has also increased,  contrary to the  expectation that
temperature   would  cause   phase  to   randomize  and   thus  reduce
$R[\theta]$. The reason is  that the $P(\theta)$ distribution at $T=0$
has   a  bimodal  (double-peaked   or  double-bunched)   form,  having
relatively large variance and  thus low $R[\theta]$. As temperature is
increased to $T=1$, this  bimodal distribution at first collapses into
a single-peaked form, the resulting sharp reduction in variance, being
responsible  for the  rise in  $R[\theta]$. With  further  increase in
temperature,  the  expected  diffusion  of  the  phase  sets  in,  and
$R[\theta]$      registers      a      gradual     reduction.

\subsubsection{System of an anharmonic oscillator}

We  consider the  system  $S$  of an  anharmonic  oscillator with  the
Hamiltonian
\begin{equation}
H_S = \hbar \omega \left( a^{\dagger} a + {1 \over 2} \right) + 
{\hbar \lambda \over 2} (a^{\dagger})^2 a^2. \label{anho}
\end{equation}
As shown in  \cite{sb06}, the  above Hamiltonian  can be
expressed in terms of the generators  of the group SU(1,1) as a result
of which the appropriate basis for it would be $|m, k\rangle$ where $m
= 0, 1, 2,... $ and $k$ equal to $1 \over 4$ or $3 \over 4$.  The case
of $k  = {1 \over  4}$ corresponds to  states with even  photon number
with the vacuum state coinciding with the vacuum state of the harmonic
oscillator, while the case of $k  = {3 \over 4}$ corresponds to states
with odd  photon number.   Using the properties  of the  SU(1,1) group
generators, the action of $H_S$ (\ref{anho}) on the basis is found to be
\begin{eqnarray}
H_S |m, k\rangle &=& 2 \hbar \left[\omega(m+k) + \lambda 
m(m+2k-1)\right] |m, k\rangle \nonumber \\ & = & E_{m_k} |m, 
k\rangle. \label{4e} 
\end{eqnarray} 
We make use of this to obtain the phase distribution of the anharmonic
oscillator system, interacting with a  squeezed thermal bath via a QND
system-bath  interaction, and starting  from the  following physically
interesting initial states: 

(A). \underline{System initially in a Kerr state}:

The initial density matrix of the system is \cite{gg94}
\begin{equation}
\rho^s(0) = |\psi_K \rangle \langle \psi_K|. \label{kerr}
\end{equation}
Here $|\psi_K \rangle$ is defined in terms of the number states as 
\begin{equation}
|\psi_K \rangle = \sum\limits_{n} q_n |n \rangle, \label{kerrfock} 
\end{equation} 
where
\begin{equation}
q_n = {\alpha^n \over \sqrt{n!}} e^{-|\alpha|^2 \over 2} e^{-
i \chi n(n-1)}. \label{qn} 
\end{equation}
In the above equations, $|n \rangle$ represents the usual number state
and  $\chi  =  {\lambda L  \over  2  v}$,  where  $\lambda$ is  as  in
Eq. (\ref{anho}), $L$ is the length of the medium and $v$ is the speed
of light in the Kerr medium  in which the interaction has taken place.
Making use of Eqs.  (\ref{qndred}), (\ref{kerr}) in Eq.  (\ref{osph}),
the phase distribution is obtained as \cite{sb06}
\begin{eqnarray}
{\cal P}(\theta)  & = &  {1 \over 2\pi}  \sum\limits_{m, n=0}^{\infty}
q_{2m}  q^{*}_{2n}  e^{i2(n-m)\theta}   e^{-2i(m-n)  \left[  \omega  +
\lambda(m+n-{1  \over  2})\right]t}  \nonumber  \\ &  &  \times  e^{4i
\hbar^2 (m-n) \left[ \omega  + \lambda(m+n-{1 \over 2} )\right] \left[
\omega(n+m+{1   \over  2})+   \lambda(n^2  +   m^2  -   {1   \over  2}
(m+n))\right]\eta(t)} \nonumber  \\ &  & \times e^{-4  \hbar^2 (m-n)^2
\left[\omega + \lambda(m+n- {1 \over 2})\right]^2 \gamma(t)} \nonumber
\\  &  &  +  {1  \over 2\pi}  \sum\limits_{m,  n=0}^{\infty}  q_{2m+1}
q^{*}_{2n+1}   e^{i2(n-   m)\theta}   e^{-2i(m-n)  \left[   \omega   +
\lambda(m+n+{1  \over 2})\right]  t}  \nonumber \\  &  & \times  e^{4i
\hbar^2 (m-  n)\left[\omega + \lambda(m+n+{1 \over  2} )\right] \left[
\omega(n+m+{3   \over  2})+   \lambda(n^2  +   m^2  +   {1   \over  2}
(m+n))\right]\eta(t)} \nonumber  \\ & & \times e^{-4  \hbar^2 (m- n)^2
\left[\omega   +  \lambda(m+n+   {1   \over  2})\right]^2   \gamma(t)}
. \label{kerrosph} 
\end{eqnarray}
The corresponding complementary number distribution is obtained, using
Eq. (\ref{osnu}), as
\begin{equation}
p(m) = |q_{2m}|^2 + |q_{2m+1}|^2, \label{kerrosnu}
\end{equation}
where $q_{2m}$, $q_{2m+1}$ can be obtained from Eq. (\ref{qn}). 

\begin{figure}
\includegraphics[width=7.0cm]{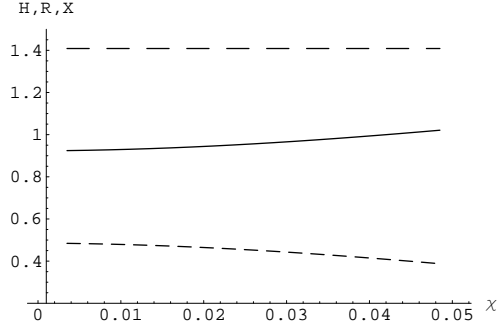}
\caption{ Number  entropy $H[m]$  (large-dashed  line),
  phase knowledge  $R[\theta]$ (small-dashed line)  and entropy excess
  $X[m,  \theta]$ (Eq.   (\ref{eq:nybial})) (bold  line) plotted  as a
  function  of  the  parameter  $\chi$ (\ref{qn})  for  an  anharmonic
  oscillator system initially  in a Kerr state and  subjected to a QND
  interaction.  The  parameters taken are  $\omega=1$, $\omega_c=100$,
  $\gamma_0=0.0025$, $|\alpha|^2=5$,  $\theta_0=0$, $\lambda=0.02$ and
  with bath squeezing parameters (\ref{eq:a}) $r=2.0$ and $a=0$ for an
  evolution time $t=0.1$ and $T=0$.}
\label{fig:comp_kerr}
\end{figure}
Using  ${\cal   P}(\theta)$  (\ref{kerrosph})   in  Eq.
(\ref{eq:phient}) to get  the phase knowledge, $p(m)$ (\ref{kerrosnu})
to get the number entropy  and using these in Eq. (\ref{eq:nybial}) we
get    the   entropy    excess   which    are   plotted    in   Figure
\ref{fig:comp_kerr}.  From the  figure , it  is evident
that  as the Kerr  parameter $\chi$  increases, phase  gets randomized
leading   to  the   fall  in   $R[\theta]$,  whereby   entropy  excess
$X[m,\theta]$ increases, since $H[m]$ remains unchanged. 
The invariance of  $H[m]$ under change in the  parameter $\chi$ can be
easily seen  by using Eq.   (\ref{qn}) in Eq.   (\ref{kerrosnu}).  The
principle of  entropy excess, Eq.  (\ref{eq:nybial}),  is clearly seen
to be satisfied. 

(B). \underline{System initially in a squeezed Kerr state}:

The initial density matrix of the system is \cite{gg94}
\begin{equation}
\rho^s(0) = |\psi_{SK} \rangle \langle \psi_{SK}|. \label{sqkerr}
\end{equation}
Here $|\psi_{SK} \rangle$ is defined in terms of the number states as
\begin{equation}
|\psi_{SK} \rangle = \sum\limits_{n} s_n |n \rangle, \label{sqkerrfock} 
\end{equation} 
where 
\begin{equation}
s_{2m} = \sum\limits_{p} q_{2p} G _{2m2p}(z), \label{s2m} 
\end{equation}
and
\begin{equation}
s_{2m+1} = \sum\limits_{p} q_{2p+1} G _{2m+12p+1}(z), 
\label{s2m1}
\end{equation}
with $z=r_1 e^{i\psi}$, and $G_{mp}(z) = \langle m | S(z) | p 
\rangle$, where $S(z)$ is the usual squeezing operator, is 
given by \cite{sm91} 
\begin{eqnarray}
G_{2m2p} & = & {(-1)^p \over p! m!} \left({(2p)! (2m)! \over 
\cosh(r_1)}\right)^{1 \over 2} \exp{\left(i(m - p)\psi \right)} 
\nonumber \\ & & \times \left({\tanh(r_1) \over 2} 
\right)^{(m+p)} F^2_1 \left[-p, -m; {1 \over 2}; -{1 \over 
(\sinh(r_1))^2}\right]. \label{Geven} 
\end{eqnarray}
Similarly, $G_{2m+12p+1}(z)$ is given by 
\begin{eqnarray}
G_{2m+12p+1} &=& {(-1)^p \over p! m!} \left({(2p+1)! (2m+1)! 
\over \cosh^3(r_1)}\right)^{1 \over 2} \exp{\left(i(m - p)\psi 
\right)} \nonumber \\ & & \times \left({\tanh(r_1) \over 2} 
\right)^{(m+p)} F^2_1 \left[-p, -m; {3 \over 2}; -{1 \over 
(\sinh(r_1))^2}\right]. \label{Godd} 
\end{eqnarray}
Here $F^2_1$ is the Gauss hypergeometric function \cite{ETBM}.  Making
use of  Eqs. (\ref{qndred}), (\ref{sqkerr}) in  Eq.  (\ref{osph}), the
phase distribution is obtained as \cite{sb06}
\begin{eqnarray}
{\cal P}(\theta) & = & {1 \over 2\pi} \sum\limits_{m, 
n=0}^{\infty} s_{2m} s^{*}_{2n} e^{i2(n-m)\theta} e^{-2i(m-
n)\left[\omega + \lambda(m+n-{1 \over 2})\right]t} \nonumber \\ 
& & \times e^{4i \hbar^2 (m-n)\left[\omega + \lambda(m+n-{1 
\over 2} )\right] \left[\omega(n+m+{1 \over 2})+ \lambda(n^2 + 
m^2 -{1 \over 2} (m+n))\right]\eta(t)} \nonumber \\ & & \times 
e^{-4 \hbar^2 (m-n)^2 \left[\omega + \lambda(m+n-{1 \over 
2})\right]^2 \gamma(t)} \nonumber \\ & & + {1 \over 2\pi} 
\sum\limits_{m, n=0}^{\infty} s_{2m+1} s^{*}_{2n+1} e^{i2(n-
m)\theta} e^{-2i(m-n)\left[\omega + \lambda(m+n+{1 \over 
2})\right]t} \nonumber \\ & & \times e^{4i \hbar^2 (m-
n)\left[\omega + \lambda(m+n+{1 \over 2} )\right] 
\left[\omega(n+m+{3 \over 2}) + \lambda(n^2 + m^2 +{1 \over 2} 
(m+n))\right]\eta(t)} \nonumber \\ & & \times e^{-4 \hbar^2 (m-
n)^2 \left[\omega + \lambda(m+n+{1 \over 2})\right]^2 
\gamma(t)} . \label{sqkerrosph} 
\end{eqnarray}
The corresponding complementary number distribution is obtained, using
Eq. (\ref{osnu}), as
\begin{equation}
p(m) = |s_{2m}|^2 + |s_{2m+1}|^2, \label{sqkerrosnu}
\end{equation}
where $s_{2m}$,  $s_{2m+1}$ can be obtained from  Eqs. (\ref{s2m}) and
(\ref{s2m1}), respectively.

\begin{figure}
\subfigure[]{\includegraphics[width=7.0cm]{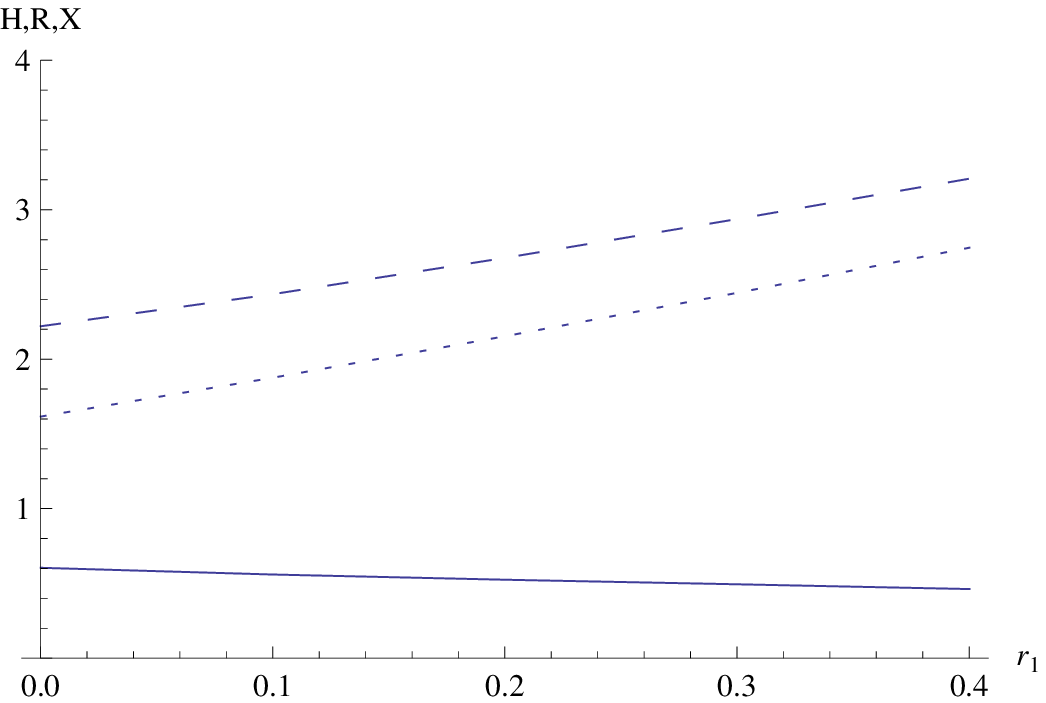}}
\subfigure[]{\includegraphics[width=7.0cm]{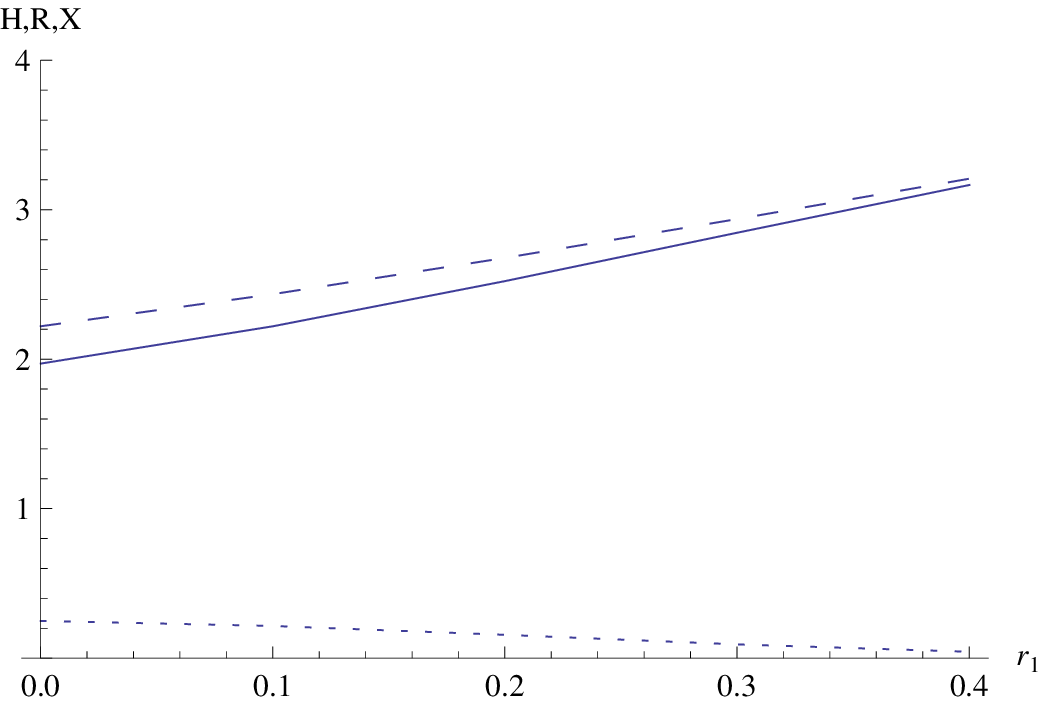}}
\caption{Plot  of  number entropy  $H[m]$  (large-dashed line),  phase
  knowledge  $R[\theta]$   (small-dashed  line)  and   entropy  excess
  $X[m,\theta]$ (bold line) for  an anharmonic oscillator initially in
  squeezed   Kerr   state,  with   respect   system  squeezing   $r_1$
  (\ref{s2m1}).  Figure  (a) represents  the  pure  state case,  while
  Figure (b) represents the system subjected to QND interaction with a
  squeezed  thermal bath  with  temperature $T=1$  and evolution  time
  $t=1$.    The  parameters   used  are   $\omega=1$,  $\omega_c=100$,
  $\gamma_0=0.1$,    $\psi=\pi/4$,    $|\alpha|^2=5$,    $\theta_0=0$,
  $\chi=0.02$,  $\lambda=0.02$,  and  with bath  squeezing  parameters
  $r=0.1$,  $a=0$.}
\label{fig:sqanharmonisk}
\end{figure}
Using  ${\cal   P}(\theta)$  (\ref{sqkerrosph})  in  Eq.
(\ref{eq:phient})    to    get    the    phase    knowledge,    $p(m)$
(\ref{sqkerrosnu})  to  get the  number  entropy  and  using these  in
Eq. (\ref{eq:nybial}) we  get the entropy excess which  are plotted in
Figures  \ref{fig:sqanharmonisk}.   In Figure  \ref{fig:sqanharmonisk}
(a),  depicting unitary  evolution, it  is seen  that  phase knowledge
almost exactly compensates for the growth of ignorance of number, as a
functions of  $r_1$, whereas , in  Figure \ref{fig:sqanharmonisk} (b),
phase knowledge  is rapidly lost,  depicting clearly the  influence of
the   environment.     The   principle   of    entropy   excess,   Eq.
(\ref{eq:nybial}),  is  clearly  seen  to  be  satisfied  for  unitary
evolution as  well as when  the anharmonic system is  interacting with
its environment. 

\subsection{Dissipative system-bath interaction}

Here  the  system-reservoir  interaction  is  such  that
$[H_{S},H_{SR}]   \neq  0$   resulting  in   decoherence   along  with
dissipation. 

(A). System of harmonic oscillator interacting with a thermal bath
resulting in a Lindblad evolution:

The initial state of the  system is a superposition of coherent states
which are $180^{\circ}$ out  of phase with respect to each
other \cite{mb92}.
\begin{equation}
|\psi \rangle = A^{1/2} (|\alpha \rangle  + e^{i\phi} |- \alpha
\rangle), \label{psi}
\end{equation}
where $\alpha = |\alpha|e^{i\phi_0}$ and 
\begin{equation}
A = \frac{1}{2}[1 + \cos(\phi) e^{-2|\alpha|^2}]^{-1}. \label{a}
\end{equation}
The state  $|\psi \rangle$ for  $\phi = 0$  would be an  even coherent
state and for $\phi = \pi$ would be an odd coherent state. The reduced
density matrix can be shown to have the following form 
\cite{vb93}:
\begin{equation}
\rho (t) = \sum\limits_{n, m=0}^{\infty} \rho_{n, m} (t) |n \rangle
\langle m|, \label{reddissos}
\end{equation}
where
\begin{eqnarray}
\rho_{n, m}(t) &=& \frac{A}{N(t) + 1} \left(\frac{e^{-\gamma_0 t/2}}
{N(t) + 1}\right)^{m + n} Q_n Q_m e^{i (n -m) \phi_0} \nonumber\\
&\times& \sum\limits_{l=0}^{\infty} \left(1 - \frac{e^{-\gamma_0 t/2}}
{N(t) + 1} \right)^l \frac{|\alpha|^{2l}}{l!} \left(1 + (-1)^{n+m} +
(-1)^l [(-1)^n e^{i \phi} + (-1)^m e^{-i \phi}] \right) \nonumber\\
&\times& F^2_1 \left[-m, -n; l + 1; 4 N(t)(N(t) + 1)
(\sinh(\gamma_0 t/2))^2\right]. \label{rhonmhodiss} 
\end{eqnarray}
Here $F^2_1$ is the Gauss hypergeometric function \cite{ETBM},
$\gamma_0$ is a parameter which depends upon the 
system-reservoir coupling strength, 
\begin{equation}
Q_n = \frac{|\alpha|^n}{\sqrt{n!}} e^{-\frac{|\alpha|^2}{2}},
\end{equation}
and, 
\begin{equation}
N(t) = N_{th} (1 - e^{- \gamma_0 t}), ~ N_{th} = \left(e^{\frac{\hbar \omega}
{k_B T}} - 1 \right)^{-1}. \label{nth}
\end{equation}
The phase distribution is given by
\begin{equation}
{\cal P}(\theta) = {1 \over 2\pi} \sum\limits_{m, 
n=0}^{\infty} \rho_{m, n} e^{i (n -m) \theta}, \label{phhodiss}
\end{equation}
where $\rho_{m, n}$ can be obtained from Eq. (\ref{rhonmhodiss}).

The corresponding complementary number distribution is obtained, using
Eq. (\ref{osnu}), as
\begin{equation}
p(m) = \rho_{m, m}(t), \label{pmhodiss}
\end{equation}
where $\rho_{m, m}$ is as in Eq. (\ref{rhonmhodiss}).

\begin{figure}
\subfigure[]{\includegraphics[width=7.0cm]{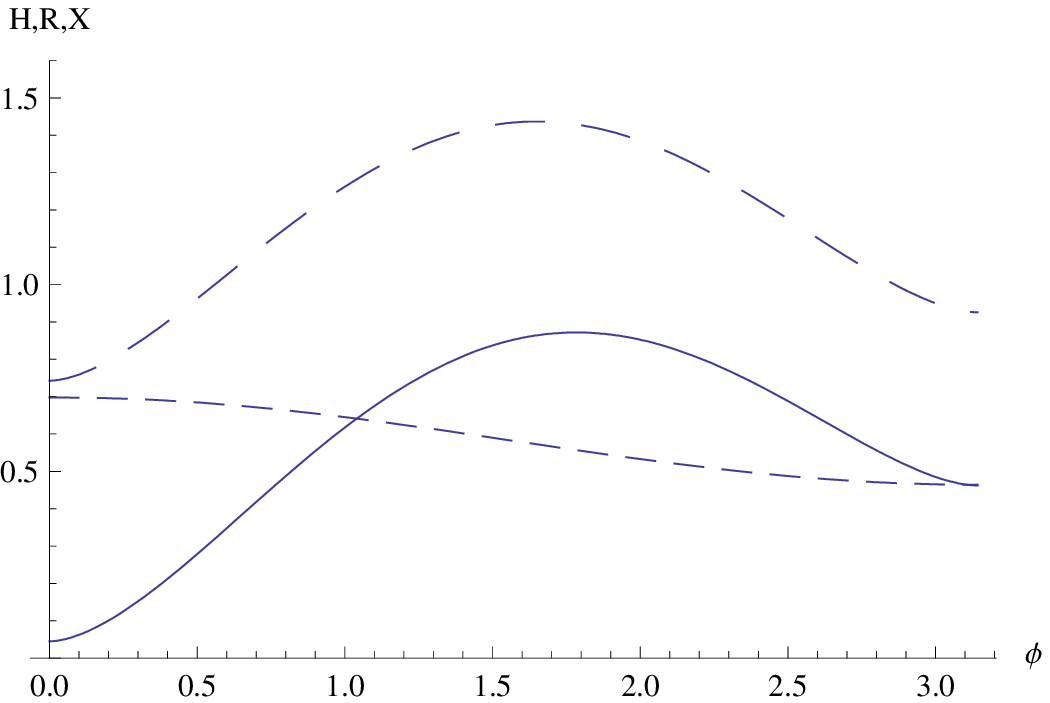}}
\subfigure[]{\includegraphics[width=7.0cm]{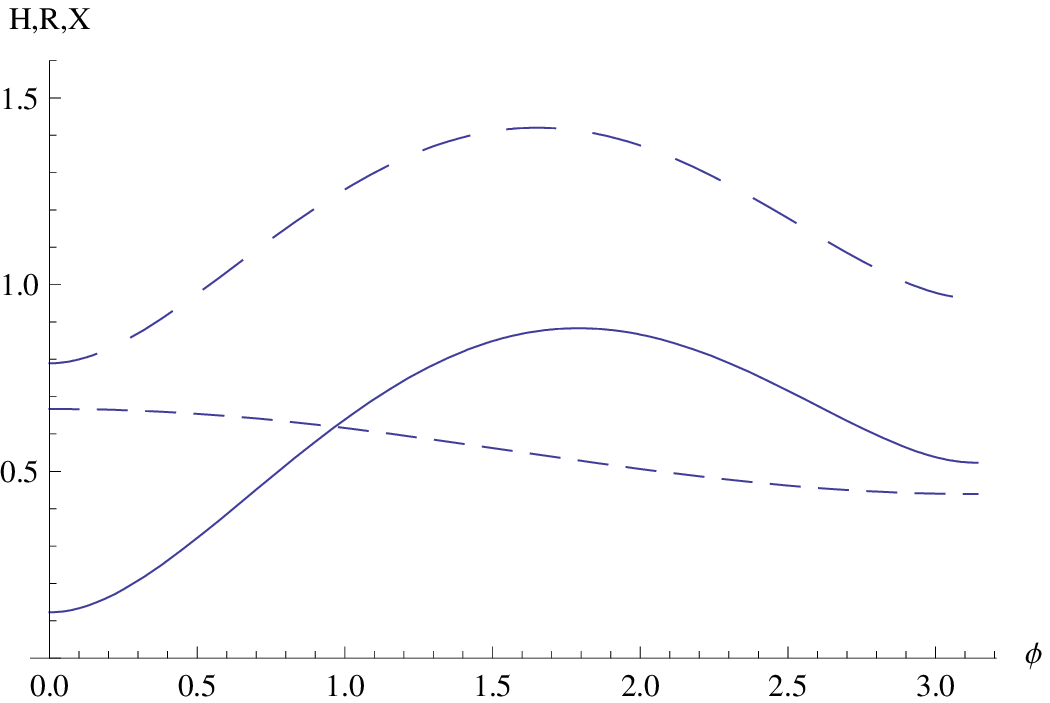}}
\caption{Plot  of  number entropy  $H[m]$  (large-dashed
  line), phase  knowledge $R[\theta]$ (small-dashed  line) and entropy
  excess  $X[m,\theta]$   (bold  line)  for   a  harmonic  oscillator,
  initially  in  a  coherent  state superposition  (\ref{psi}),  as  a
  function of  the state parameter  $\phi$ (\ref{psi}).  
  Figure (a) pertains  to the pure state case.   Figure (b) represents
  the  system   subjected  to  a  dissipative   interaction  with  the
  environment  for an  evolution time  $t=0.1$ and  temperature $T=2$.
  The    parameters    used    are    $\omega=1$,    $\gamma_0=0.025$,
  $|\alpha|^2=2$,  $\phi_0$  (\ref{psi})  $=0$.}
\label{fig:ho_entxs_t0p1_T1.eps}
\end{figure}
Using  ${\cal   P}(\theta)$  (\ref{phhodiss})   in  Eq.
(\ref{eq:phient}) to get  the phase knowledge, $p(m)$ (\ref{pmhodiss})
to get the number entropy  and using these in Eq. (\ref{eq:nybial}) we
get   the    entropy   excess    which   are   plotted    in   Figures
\ref{fig:ho_entxs_t0p1_T1.eps}.               From              Figure
\ref{fig:ho_entxs_t0p1_T1.eps}  (a), pertaining to  unitary evolution,
we note that in the even cat (coherent) state ($\phi=0$), ignorance of
number approximately  equals phase knowledge,  whereas in the  odd cat
(coherent) state  ($\phi=\pi$) the former  significantly outweighs the
latter. This thus provides a complementaristic characterization of the
even and odd cat states. The Figure \ref{fig:ho_entxs_t0p1_T1.eps} (b)
shows that the  effect of the dissipative environment  causes phase to
become  randomized, leading  to  an increased  entropy  excess at  all
$\phi$   (\ref{psi}).    The   principle   of  entropy   excess,   Eq.
(\ref{eq:nybial}), is  clearly seen to be satisfied,  for both unitary
as well as dissipative evolution. 

(B). System of anharmonic oscillator weakly interacting with a thermal bath:

The  total Hamiltonian depicting  a third-order  non-linear oscillator
coupled  to a  reservoir  of oscillators  \cite{jp84},  assumed to  be
initially in a thermal state, is
\begin{equation} 
H = \hbar \left[\omega(a^{\dagger}a + \frac{1}{2}) + \kappa a^{\dagger 2} a^2
+ \sum\limits_j \omega_j (b^{\dagger}_j b_j + \frac{1}{2}) + \sum\limits_j
(\kappa_j b_j a^{\dagger} + \kappa^*_j b^{\dagger}_j a) \right]. \label{anhar}
\end{equation}
The reduced density matrix  of the anharmonic oscillator, 
starting from  the initial  coherent state $|\xi  (0) \rangle$  $=$ $|
|\xi(0)| e^{i  \phi_0} \rangle$, can be  solved and made
use of to obtain the phase distribution
\begin{eqnarray}
{\cal P} (\theta) &=& {1 \over 2\pi} \sum\limits_{m, 
n=0}^{\infty} \rho_{m, n} e^{i (n -m) \theta} \nonumber\\
&=& \frac{1}{2} \sum\limits_{m, 
n=0}^{\infty} (m! n!)^{1/2} f_{n, m} (t) e^{i (n -m) \theta}. \label{phanhar} 
\end{eqnarray}
Here 
\begin{eqnarray}
f_{m, n} (t) &=& \exp\left([-2i\kappa (m-n) + \frac{\gamma_0}{2}]t\right) 
\left(E_{m-n}(t)\right)^{m+n+1} \sum\limits_{l=0}^{\infty} \frac{1}{l!}
\left[\frac{N_{th} + 1}{N_{th}} g_{m-n}(t)\right]^l \frac{(m+l)! (n+l)!}
{m! n!} \nonumber\\ &\times& f_{m+l, n+l}(0)  F^2_1 \left
[-m, -n; l + 1; \frac{4 N_{th}(N_{th} + 1)}{\Delta^2}
(\sinh(\gamma_0 \Delta t/2))^2\right]. \label{rhomnanhodiss} 
\end{eqnarray}
In the above equation, $f_{m+l, n+l}(0)$ contains 
information about the initial state of the system and for the 
initial coherent state $|\xi (0) \rangle$ $=$ $| |\xi(0)| 
e^{i \phi_0} \rangle$ is given by
\begin{equation}
f_{m,n}(0)= \frac{1}{\pi}\frac{\xi^{m*}(0)}{m!}\frac{\xi^n}{n!}.
\end{equation} 
$F^2_1$ is the Gauss hypergeometric function \cite{ETBM},
$\gamma_0$ is a parameter which depends upon the 
system-reservoir coupling strength and 
$N_{th}$ is as defined above. Also
\begin{equation}
E_{m-n}(t) = \frac{\Delta}{\Omega \sinh(\gamma_0 \Delta t/2) +
\Delta \cosh(\gamma_0 \Delta t/2)}, \label{emn}
\end{equation}
\begin{equation}
g_{m-n}(t) = \frac{2 N_{th}}{\Omega  +
\Delta \coth(\gamma_0 \Delta t/2)}, \label{gmn}
\end{equation}
\begin{equation}
\Omega \equiv \Omega_{m-n} = 1 + 2 N_{th} - i \frac{2 \kappa}{\gamma_0}
(m-n), \label{omegamn}
\end{equation}
and
\begin{equation}
\Delta \equiv \Delta_{m-n} = \left[\Omega^2 - 4 N_{th} (N_{th} + 1)
\right]^{1/2}.  \label{deltamn}
\end{equation}

The corresponding complementary number distribution is obtained, using
Eq. (\ref{osnu}), as
\begin{equation}
p(m) = \pi n! f_{m, m}(t), \label{pmanhodiss}
\end{equation}
where $f_{m, m}$ can be obtained from Eq. (\ref{rhomnanhodiss}).

\begin{figure}
\includegraphics[width=7.0cm]{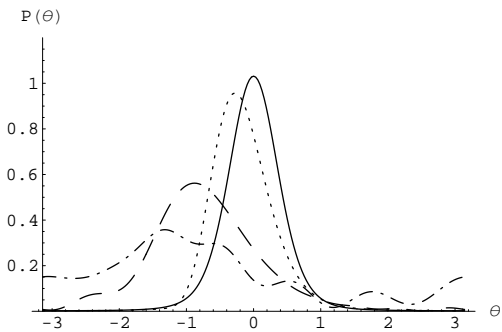}
\caption{Phase distribution  ${\cal P}(\theta)$ as a function of
  $\theta$ for  the dissipative  anharmonic oscillator initially  in a
  coherent state.   The bold curve represents  unitary evolution while
  the  other curves  represent temperature  $T=0$ and  evolution times
  $t=1,  5$  and $10$  for  the  dotted,  large-dashed and  dot-dashed
  curves,   respectively.   The   parameters   used  are   $\omega=1$,
  $|\xi(0)|^2=2$,  $\phi_0=0$ ($|\xi(0)|$,  $\phi_0$  are the  initial
  state     parameters),     $\gamma_0=0.01$     and     $\kappa=0.05$
  (\ref{anhar}).}
\label{fig:figur5anh}
\end{figure}

\begin{figure}
\subfigure[]{\includegraphics[width=7.0cm]{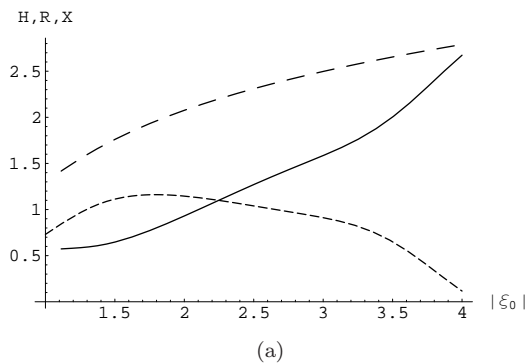}}
\caption{Number  entropy $H[m]$  (large-dashed line),  phase knowledge
  $R[\theta]$  (small-dashed line)  and  entropy excess  $X[m,\theta]$
  (bold  line) for  an anharmonic  oscillator system,  initially  in a
  coherent state,  with respect  to initial state  parameter $|\xi_0|$
  ($=|\xi(0)|$),  subjected  to a  dissipative  interaction where  the
  parameters  are as  in  the  above figure.   The  figures depict  an
  evolution time $t=2$ and temperature $T=0$. It can be shown that, in
  contrast   to    the   corresponding   harmonic    oscillator   case
  (Fig. \ref{fig:inicoh}(a)),  increase in average  number ($|\xi_0|$)
  is not  accompanied by a corresponding increase  in phase knowledge.
}
\label{fig:anho_srx_t2_T0}
\end{figure}
 The   phase    distribution   ${\cal    P}   (\theta)$
(\ref{phanhar}) is plotted in Figure \ref{fig:figur5anh} from which we
see that  with increase  in time, phase  gets randomized  resulting in
phase diffusion.  Using this ${\cal P}(\theta)$ (\ref{phanhar}) in Eq.
(\ref{eq:phient})    to    get    the    phase    knowledge,    $p(m)$
(\ref{pmanhodiss})  to  get the  number  entropy  and  using these  in
Eq. (\ref{eq:nybial}) we  get the entropy excess which  are plotted in
Figure  \ref{fig:anho_srx_t2_T0}.   The  effect  of   the  dissipative
interaction is  seen to manifest in the  increased phase randomization
and  entropy excess with  increase in  the state  parameter $|\xi_0|$.
The principle  of entropy  excess, Eq.  (\ref{eq:nybial}),  is, again,
clearly seen to be satisfied. 

\section{Atomic System}

Here  we  study  entropy  excess  (\ref{eq:nybial})  for  number-phase
complementarity in  (finite-level) atomic systems,  briefly revisiting
results  obtained  from the  perspective  of  an  upper bound  on  the
knowledge-sum  of complementary  variables in  Ref.   \cite{rs07}.  An
interesting  generalization  of  the  knowledge-sum  of  complementary
variables  could be  made, in  the context  of  quantum communication,
using  the information exclusion  relations developed  in \cite{mh95}.
As pointed  out earlier, the knowledge-sum approach  cannot be applied
to  infinite dimensional  systems,  whereas the  principle of  entropy
excess  can be  applied  to  finite as  well  as infinite  dimensional
systems,  making it  a more  flexible tool  for  studying number-phase
complementarity in a host of systems.

As an  application of  the principle, it  is appropriate to  study the
effect of  noise. This  we do for  noise from both  non-dissipative as
well as  dissipative interactions  of the atomic  system $S$  with its
environment,  which is  modelled  as a  bath  of harmonic  oscillators
starting in a squeezed thermal state \cite{bg06,gp,sqgen}.  In Section
\ref{sec:openphase}  we  consider  the  effect of  the  phase  damping
channel,  which   is  the   information  theoretic  analogue   of  the
non-dissipative  open system effect  \cite{bg06,gp}, while  in Section
\ref{sec:opengen} we  consider the effect of  the squeezed generalized
amplitude damping channel which  is the information theoretic analogue
of the dissipative open system effect \cite{gp,sqgen}.

\subsection{The principle of entropy excess in atomic systems}

For a (noiseless) two-level (spin-1/2) system, the plot of entropy  $H[m]$ 
for all atomic  coherent states is given by  the large-dashed curve in
Figure \ref{fig:atomXS}(a).  The equatorial states on the Bloch sphere,
corresponding  to   $\alpha'=\pi/2$,  are  the   maximum
knowledge  state  (MXK)   states  of  $\phi$,  and  are
precisely  equivalent  to  the  minimum  knowledge  state
(MNK) states of  $m$ (characterized by $H[m]=1$), as can
be  seen from comparing  the large-dashed  and small-dashed  curves in
the Figure.  Thus number  and phase share with MUBs the
reciprocal  property  that  maximum   knowledge  of  one  of  them  is
simultaneous  with minimal knowledge  of the  other, but  differs from
MUBs in  that the  maximum possible knowledge  of $\phi$ is  less than
$\log(d) = 1$ bit, essentially on account of its POVM nature.

Two  variables form  a  {\it quasi-MUB}  if  any MXK  state of  either
variable is an MNK state of  the other, where the knowledge of the MXK
state may  be less  than $\log  d$ bits.  Thus,  $J_z$ and  $\phi$ are
quasi-MUB's (but not MUB's).

From  the  dot-dashed  curve  in the  Figures  (\ref{fig:atomXS}),  we
numerically find an expression of the uncertainty principle to be
\begin{equation}
\label{eq:1bit}
X[m,\phi] \equiv H[m] - R[\phi] \ge 0
\end{equation}
for  all   pure  states  in   ${\bf  C}^2$,  in  conformity   with  Eq.
(\ref{eq:ra})  and  hence  also in  agreement  with  the
principle of entropy  excess (\ref{eq:nybial}). As $\phi$
is a  POVM but $m$ represents a regular Hermitian observable, 
in general  $X[m,\phi] \ne X[\phi,m]$.  The  inequality is
saturated only for the Wigner-Dicke states (as seen from the
dot-dashed curve in the Figure), when $H[m]$ and $R[\phi]$ identically
vanish.

As  an   expression  of   the  uncertainty  principle,   the  relation
(\ref{eq:1bit}) still leaves some  room for improvement.  First, it is
not a  tight bound.  In  particular, for equatorial states  it permits
$R[\phi]$ to  be as high as  1, whereas as seen  from the 
small-dashed   curve  in  Figure   \ref{fig:atomXS},  the
maximum value of $R[\phi]$, which is $r_\phi \approx 0.245$.

Following  Ref. \cite{rs07},  one way  to address  this problem  is to
modify (\ref{eq:1bit}) to the inequality
\begin{equation}
\label{eq:2bit}
X^\mu[m,\phi] \equiv H[m] - \mu R[\phi] \ge 0
\end{equation}
for  all pure states  in ${\bf  C}^2$, where  parameter $\mu$ $(>0)$ is
chosen to be the largest value such that inequality (\ref{eq:2bit}) is
satified over  all state space.  Through a numerical search,  we found
that $\mu \approx  4.085$ for dimension $d=2$ and  $\mu \approx 1.973$
for $d=4$. From the concavity of $H[m]$ and the convexity of
$R[\phi]$, it follows that Eq. (\ref{eq:2bit}) holds for any mixed
state.
The small-dashed  and dotted  curves are, respectively,  $R[\phi]$ and
$\mu R[\phi]$.  
Comparing their corresponding curves in the  Figure, we note the tighter
bound imposed by $X_\mu[m,\phi]$ than $X[m,\phi]$.

\subsection{Application to the phase damping channel \label{sec:openphase}}

The  `number'  and  phase  distributions  for  a  qubit,  
$H_S=\frac{\hbar\omega}{2}\sigma_z$,  starting from  an
atomic   coherent   state  $|\alpha^\prime,\beta^\prime\rangle$,   and
subjected to  a phase  damping channel due  to its interaction  with a
squeezed thermal bath, are \cite{sb06,bg06}
\begin{eqnarray}
p(m) &=& \left( \begin{array}{c} 2j \\ j+m\end{array}\right) 
(\sin(\alpha^{\prime}/2))^{2(j+m)}
(\cos(\alpha^{\prime}/2))^{2(j-m)} \nonumber \\
{\cal P}(\phi) &=& {1 \over 2 \pi}\left[1 + {\pi \over 4} 
\sin(\alpha^{\prime}) \cos(\beta^{\prime} + \omega t - \phi) e^{- 
(\hbar \omega)^2 \gamma(t)}\right].  
\label{eq:atomcohqnd}
\end{eqnarray}

\begin{figure}
\subfigure[]{\includegraphics[width=7.0cm]{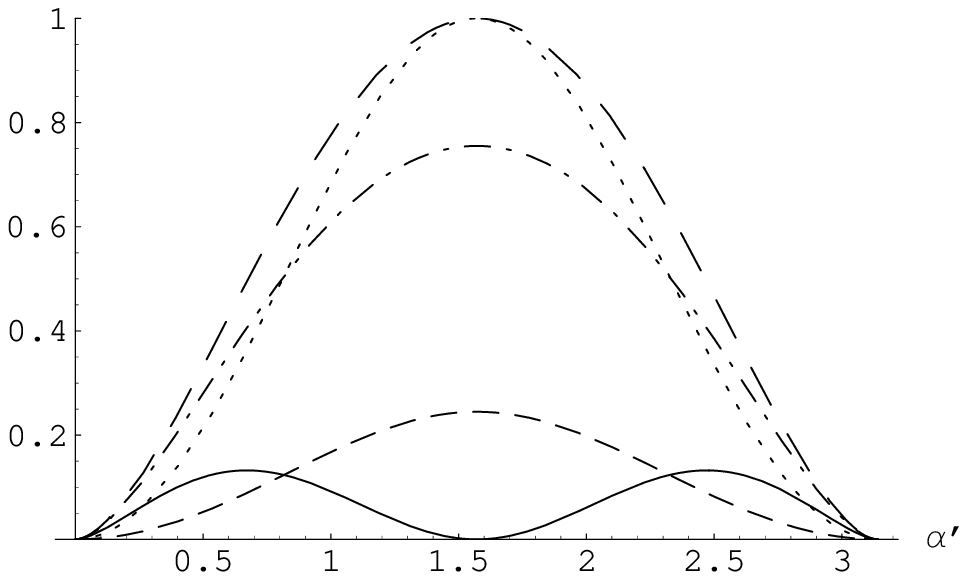}}
\subfigure[]{\includegraphics[width=7.0cm]{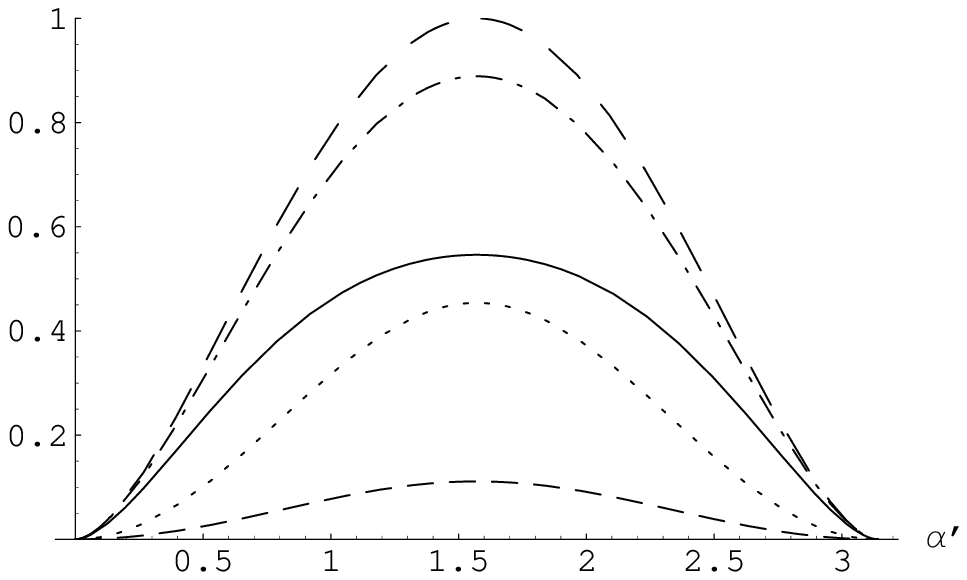}}
\caption{Entropy excess  of a two level system subjected
  to   QND  interaction   starting   in  an   atomic  coherent   state
  $|\alpha^\prime,\beta^\prime\rangle$,     as    a     function    of
  $\alpha^\prime$, with  $\beta^\prime=0.0$.  The large-dashed (resp.,
  small-dashed)  line  represents   $H[m]$  (resp.,  $R[\phi]$).   The
  dotted-curve  represents  $\mu  R[\phi]$ (where  $\mu=4.085$).   The
  solid  (resp.,  dot-dashed)  curve  represents  the  entropy  excess
  $X_\mu$ (resp.  $X$); (a) depicts  the noiseless case .  There is no
  $\beta$-dependence;  (b) depicts  the case  of QND  interaction. The
  parameters   used  are   $\omega=1.0$,   $\omega_c=100$,  $\gamma_0$
  (\ref{gamma0})  $=0.025$,  bath  squeezing  parameters  (\ref{eq:a})
  $r=0.5$ and $a=0$. The plots in the figure (b) are for a temperature
  $T=10$ and an evolution time  $t=1$.  We note  the
  symmetry  in the  figure  about $\alpha^\prime=\pi/2$.  In (a),  the
  points where $X_\mu=0$, namely  the polar and the equatorial states,
  represent the  coherent state.   If complementarity is  expressed in
  terms  of  knowledge sum  \cite{rs07},  these  states correspond  to
  maximum knowledge states.}
\label{fig:atomXS}
\end{figure}

For  completeness, the  function  $\gamma(t)$ appearing  in the  above
equation is  given in Appendix A.   We note the  symmetry preserved in
Figures  (\ref{fig:atomXS}) (a)  and (b),  about  $\alpha'=\pi/2$, the
equatorial  states.  In  the case  of  $R[\phi]$, this  is because  of
symmetry of $\sin(\alpha^\prime)$, as in Eq. (\ref{eq:atomcohqnd}) for
${\cal P}(\phi)$  about $\pi/2$,  whereas in the  case of  $H[m]$, the
symmetry  comes  about  because  the $\cos(\cdot)$  and  $\sin(\cdot)$
functions, in Eq. (\ref{eq:atomcohqnd})  for $p(m)$, appear only as an
even  power.  For  QND interaction  $p(m)$ is  time-invariant, whereas
${\cal P}(\phi)$ evolves in a way that does not affect this symmetry.

Figure \ref{fig:atomXS}(b)  depicts the effect of  phase damping noise
on  the   number  entropy  $H[m]$   (obtained  by  using   the  number
distribution $p(m)$  (\ref{eq:atomcohqnd})), phase knowledge $R[\phi]$
(obtained   by   using  the   phase   distribution  ${\cal   P}(\phi)$
(\ref{eq:atomcohqnd}))  ,  $\mu R[\phi]$,  $X[m,\phi]$  (by using  Eq.
(\ref{eq:nybial}))  and   $X_\mu[m,\phi]$.   Comparing  it   with  the
noiseless case, as in  Figure \ref{fig:atomXS}(a), we find that $H[m]$
remains  invariant  because  $p(m)$  is  not affected  when  a  system
undergoes  a QND  interaction, but  there  is an  increase in  $X_\mu$
because of phase randomization with time.

\subsection{Application to the 
squeezed generalized amplitude damping channel 
\label{sec:opengen}}
The  `number' and  phase distributions  for a  qubit starting  from an
atomic   coherent   state  $|\alpha^\prime,\beta^\prime\rangle$,   and
subjected  to   a  squeezed  generalized   amplitude  damping  channel
\cite{sqgen} due to its interaction  with a squeezed thermal bath, are
\cite{sr07},
\begin{equation}
p(m=1/2,t) 
= \frac{1}{2}\left[\left(1-\frac{\gamma_0}{\gamma^\beta}\right)
+ \left(1+\frac{\gamma_0}{\gamma^\beta}\right)e^{-\gamma^\beta t}\right]
\sin^2(\alpha^\prime/2) +
\frac{\gamma_-}{\gamma^\beta}\left(1-e^{-\gamma^\beta t}\right)
\cos^2(\alpha^\prime/2), 
\label{eq:pmc}
\end{equation}
and
\begin{eqnarray}
{\cal  P}(\phi) &=&  \frac{1}{2  \pi} \left[1  + \frac{\pi}{4  \alpha}
\sin(\alpha^{\prime})   \Big\{\alpha    \cosh(\alpha   t)\cos(\phi   -
\beta^{\prime}) +  \omega \sinh(\alpha t)  \sin(\phi - \beta^{\prime})
\right. \nonumber\\ & & \left.  - \gamma_0 \chi \sinh(\alpha t) \cos(\Phi
+  \beta^{\prime}  +   \phi)  \Big\}  e^{-\frac{\gamma^{\beta}  t}{2}}
\right].  \label{phasesqdiss}
\end{eqnarray}
A  derivation of  Eqs. (\ref{eq:pmc})  and (\ref{phasesqdiss})  can be
found in Ref. \cite{sr07}.  For completeness, the parameters appearing
in these equations are given in Appendix \ref{secap:disi}.

Figures \ref{fig:atomXSdis}(a)  and (b) depict the  effect of squeezed
generalized amplitude  damping noise on the functions  depicted in the
noiseless case  of Figure  \ref{fig:atomXS}(a), without and  with bath
squeezing, respectively.   Comparing them with the  noiseless case, we
find as expected  that noise impairs both number  and phase knowledge.
If the  dependence on $\beta^\prime$ is taken  into consideration (cf.
Ref. \cite{sr07}), it  can be shown that squeezing  has the beneficial
effect of relatively improving  phase knowledge for certain regimes of
the parameter space,  and impairing them in others.  This property can
be  shown  to improve  the  classical  channel capacity  \cite{sqgen}.
Further, bath squeezing can be  shown to render $R[\phi]$ dependent on
$\beta^{\prime}$.    On  the   other   hand,  it   follows  from   Eq.
(\ref{eq:pmc}) that  $R[m]$ is independent of  $\beta^\prime$, so that
$X[m,\phi]$ is  dependent on $\beta^\prime$.  This  stands in contrast
to  that  of  the  {\it  phase  damping  channel},  where  inspite  of
squeezing,  $X[m,\phi]$ remains  independent of  $\beta^{\prime}$ and,
furthermore, squeezing  impairs knowledge of $\phi$ in  all regimes of
the parameter space.

\begin{figure}
\subfigure[]{\includegraphics[width=7.0cm]{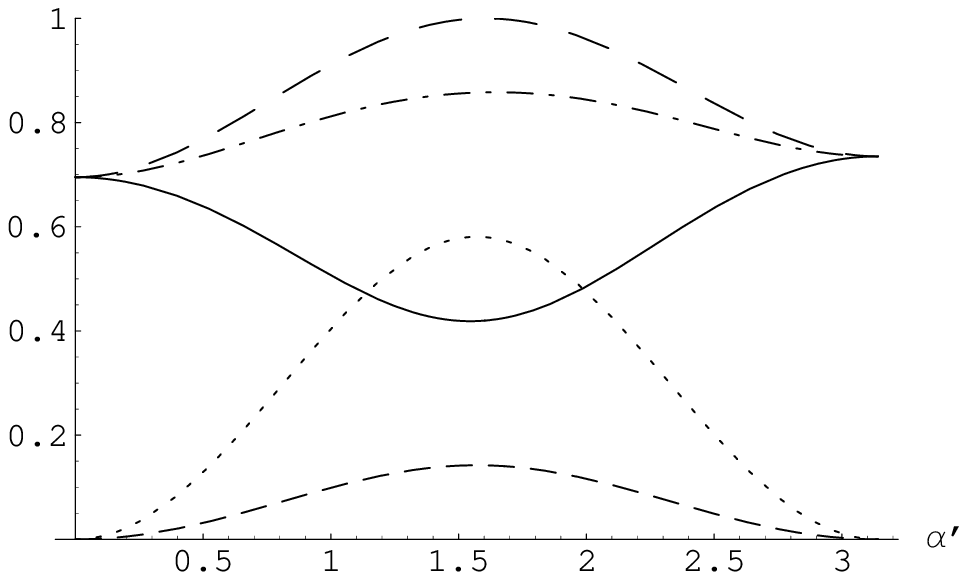}}
\subfigure[]{\includegraphics[width=7.0cm]{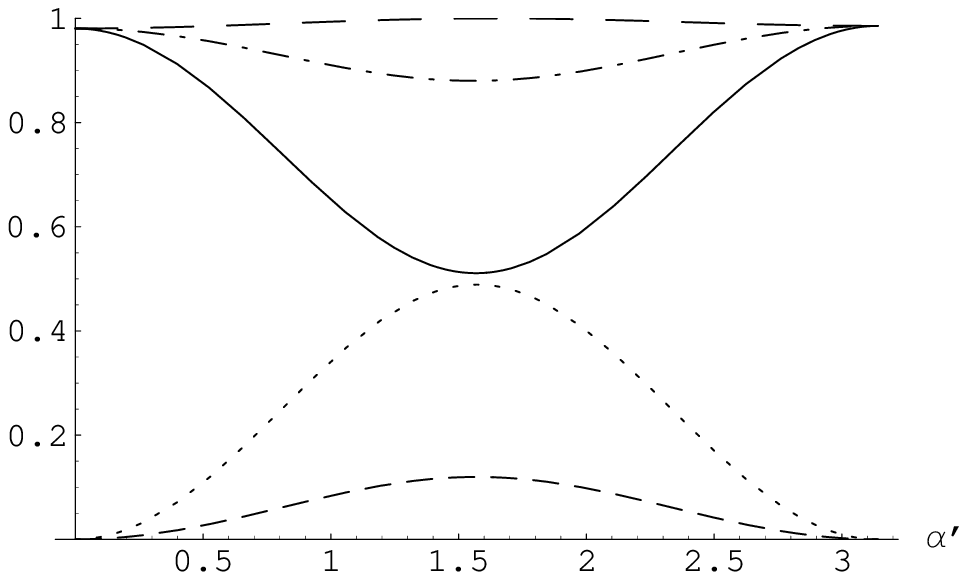}}
\caption{Entropy  excess  of  a   two  level  system  subjected  to  a
  dissipative  interaction  starting   in  an  atomic  coherent  state
  $|\alpha^\prime,\beta^\prime\rangle$,     as    a     function    of
  $\alpha^\prime$, with  $\beta^\prime=0.0$.  The large-dashed (resp.,
  small-dashed)  line  represents   $H[m]$  (resp.,  $R[\phi]$).   The
  dotted-curve   represents   $\mu   R[\phi]$.   The   solid   (resp.,
  dot-dashed) curve represents the entropy excess $X_\mu$ (resp. $X$).
  Here $\omega=1.0$,  $\omega_c=100$, $\Phi$ (\ref{eq:M})$  =\pi / 8$,
  $\gamma_0=0.025$,  with  the evolution  time  $t=1$ and  temperature
  $T=10$:  (a)  bath  squeezing  parameter $r$  (\ref{eq:M})=$0$;  (b)
  $r=1$.  Comparison of (b) with (a) shows that squeezing impairs both
  number and  phase knowledge, leading  to an increase in  the entropy
  excess $X_\mu$ (and $X$).}
\label{fig:atomXSdis}
\end{figure}

\section{Discussions and Conclusions}

In  this paper, we  have recast  the number-phase  complementarity for
finite dimensional atomic as  well as infinite dimensional oscillator,
discrete (Hermitian) as well as continuous (positive operator) valued,
systems as  a lower  bound on an  entropic measure called  the entropy
excess.  For   maximally  complementary  systems,  the   bound  is  0,
independent  of the  system  dimension.  This is  in  contrast to  the
conventional entropy sum  principle, which has a lower  bound of $\log
d$.  To tighten  the constraint imposed by the  bound on $R[\phi]$, we
replace  this quantity  by $\mu  R[\phi]$, where  $\mu$ is  a positive
number  with values (approximately)  4, 2  and 1  for two-,  four- and
infinite-dimensional  systems.  Thus  dimensional  dependence  of  the
inequality  enters  indirectly  through   the  form  $\mu  =  \mu(d)$.
Encouraged  by the above  numerical-analytical pattern,  we conjecture
that as  the system  dimension increases from  two to  infinity, $\mu$
falls monotonically from about 4 to 1.

In this  work, we have made  precise the sense in  which the variables
$p(m)$ and $P(\phi)$ may be  thought of as or differ from conventional
complementary variables  \cite{as96}.  There are  two main differences
as  follows. First: whereas  states of  maximum number  knowledge (the
eigenstates  of the  number operator)  have the  maximum  knowledge of
$\log d$ bits, the maximum phase knowledge states have less than $\log
d$ bits, phase  being a POVM. This was  the motivation for introducing
the weight  quantity $\mu$.  Second:  even more remarkably,  states of
maximum  phase   knowledge  do  not  correspond   to  equal  amplitude
superpositions of  number states. In other words,  the unbiasedness is
not  mutual,  but one-way,  a  situation  we  characterize as  one-way
unbiased bases \cite{rs07}.

In the  second aspect of  our work, the  above analysis is  applied to
physically relevant  initial conditions of  the system for  unitary as
well as  non-unitary evolution, due  to the interaction of  the system
with  its environment.  The  system-reservoir interactions  are chosen
such that both dephasing  (decoherence without dissipation) as well as
dissipative  (decoherence  with  dissipation)  effects on  the  system
evolution are studied.

Some interesting  features seen were, for e.g.,  a harmonic oscillator
starting out  from an initial superposition of  coherent (cat) states.
The  entropy  excess principle  was  seen  to  provide an  interesting
complementaristic characterization of the even and the odd cat states,
in that the excess is almost  zero for the even state, indicating that
ignorance of number approximately  equals phase knowledge while in the
odd  state, the  entropy excess  is finite  indicating that  there the
ignorance of number significantly outweighs knowledge of phase.

Our   entropy-based  formalism  can   modify  current   approaches  to
number-phase  complementarity: e.g.,  one  can  study   complementarity  in
conjunction with  such phenomena as  nonlinearity induced coherences
and atomic squeezing in an effectively finite-level atomic system.  In
the   conventional  approach,   complementarity  can   be  graphically
demonstrated  by the  constrasting behavior  of the  number  and phase
distributions (eg., Figs 1 and 2 of Ref. \cite{as96}).

As  a  concrete application  to  a  finite  dimensional system  in  an
experimental  scenario, we consider  the energy  manifold of  the four
levels of  (for instance)  $^{85}Rb$ atom. This  is first mapped  to a
pseudo-spin system of spin 3/2  while the effect of selection rules of
atomic      transitions       in      $^{85}Rb$      is      preserved
\cite{arch}.  Complementarity  can then  be  studied using  (entropic)
knowledge of  the number  and phase variables  as a function  of laser
detuning  and {\it vis-\`a-vis}  atomic phenomena  like  coherent population
trapping (CPT) or electromagnetically induced transparency (EIT).  For
example,   simulations  indicate  an   increase  in   phase  knowledge
accompanying   the  formation   of   the  CPT   state.   Noting   that
$\langle\theta,\phi|   J_-|\theta,\phi\rangle = j\sin\theta e^{i\phi}$,   
where  $J_-   \equiv J_x-J_y$, one can detect $\phi$ in a practical, 
interferometric set-up by applying $J_-$  to one of the two  interferometric arms 
implemented in an atom-laser system. With appropriate adjustments, $\phi$
will then manifest as a phase shift in the interference pattern.


\appendix

\section{Some expressions pertaining to the phase damping channel
\label{secap:qnd}}

For the case of an Ohmic bath with spectral density
\begin{equation}
I(\omega) = {\gamma_0 \over \pi} \omega e^{-\omega/\omega_c}, 
\label{gamma0} 
\end{equation}
where $\gamma_0$ and $\omega_c$ are two bath parameters characterizing
the quantum noise, it can shown that using Eq.  (\ref{gamma0}) one can
obtain \cite{bg06}
\begin{equation}
\eta (t) = -{\gamma_0 \over \pi} \tan^{-1} (\omega_c t), 
\label{2l} 
\end{equation}
and 
\begin{equation}
\gamma (t)  =  {\gamma_0 \over 2\pi} \cosh (2r) \ln 
(1+\omega^2_c t^2) - {\gamma_0 \over 4\pi} \sinh (2r) \ln 
\left[ {\left( 1+4\omega^2_c(t-a)^2\right) \over \left( 1+ 
\omega^2_c (t-2a)^2 \right)^2} \right]  - 
{\gamma_0 \over 4\pi} \sinh (2r) \ln (1+4a^2\omega^2_c) , 
\label{2.7} 
\end{equation}
in the $T  = 0$ limit, where the resulting  integrals are defined only
for $t > 2a$.  In the high  $T$ limit, $\gamma (t)$ can be shown to be
\cite{sb06}
\begin{eqnarray} 
\gamma (t) & = & {\gamma_0 k_BT \over \pi \hbar \omega_c} \cosh 
(2r) \left[ 2\omega_c t \tan^{-1} (\omega_c t) + \ln \left( {1 
\over 1+\omega^2_c t^2} \right) \right] - 
{\gamma_0 k_BT \over 2\pi \hbar \omega_c} \sinh (2r) \Bigg[ 
4\omega_c (t-a) \tan^{-1} \left( 2\omega_c (t-a) \right) 
\nonumber \\ & -& 4\omega_c (t-2a) \tan^{-1} \left( \omega_c 
(t-2a) \right) + 4a\omega_c \tan^{-1} \left( 2a\omega_c \right) 
+ \ln \left( {\left[ 1+\omega^2_c (t-2a)^2 
\right]^2 \over \left[ 1+4\omega^2_c (t-a)^2 \right]} \right) + 
\ln \left( {1 \over 1+4a^2\omega^2_c} \right) \Bigg] , 
\label{eq:gamma} 
\end{eqnarray} 
where, again, the resulting integrals are defined
for $t  > 2a$. Here we have for  simplicity taken the squeezed
bath parameters as
\begin{eqnarray} 
\cosh \left( 2r(\omega) \right) & = & \cosh (2r),~~ \sinh 
\left( 2r (\omega) \right) = \sinh (2r), \nonumber\\ \Phi 
(\omega) & = & a\omega, \label{eq:a} 
\end{eqnarray} 
where $a$ is a constant depending upon the squeezed bath.  The results
pertaining to a thermal bath  can be obtained from the above equations
by setting the squeezing parameters $r$ and $\Phi$ to zero.

\section{Some expressions pertaining to the squeezed generalized
amplitude damping channel
\label{secap:disi}} 

Here the reduced  dynamics of the two level  atomic system interacting
with a  squeezed thermal  bath under a  weak Born-Markov  and rotating
wave  approximation is  studied.  This  implies that  here  the system
interacts with its environment via a non-QND interaction, i.e., $[H_S,
H_{SR}]  \ne 0$  such that  along with  a loss  in  phase information,
energy  dissipation also takes  place.  

The parameter $\alpha$ (Eq. (\ref{phasesqdiss})) is given by
\begin{equation}
\alpha = \sqrt{\gamma^2_0 |M|^2 - \omega^2}. \label{3n}
\end{equation}
Further
\begin{equation}
\gamma^{\beta} = \gamma_0 (2 N + 1),
\label{eq:gammabeta}
\end{equation}
and 
\begin{equation}
\gamma_- = \gamma_0 N,
\label{eq:gammaminus}
\end{equation}
where 
\begin{equation}
N = N_{\rm th}(\cosh^2(r) + \sinh^2(r)) + \sinh^2(r), \label{4a7} 
\end{equation}
\begin{equation}
M = -\frac{1}{2} \sinh(2r) e^{i\Phi} (2 N_{\rm th} + 1)
\equiv \chi e^{i\Phi}, \label{eq:M}
\end{equation}
and
\begin{equation}
N_{\rm th} = {1 \over e^{{\hbar \omega \over k_B T}} - 1}. \label{4a9}
\end{equation}
Here  $N_{\rm th}$  is the  Planck distribution  giving the  number of
thermal  photons  at  the  frequency  $\omega$  and  $r$,  $\Phi$  are
squeezing parameters.   The analogous case  of a thermal  bath without
squeezing can be obtained from  the above expressions by setting these
squeezing parameters to zero.  $\gamma_0$ is  a constant  typically
denoting the system-environment coupling strength.

\end{document}